\documentclass[acmsmall,screen]{acmart}
\setcopyright{none}            %% For review submission

\title[Optimal Stateless Model Checking under the Release-Acquire Semantics]{Optimal Stateless Model Checking under the Release-Acquire Semantics}

\usepackage{wrapfig,mathtools}
\usepackage[ruled,vlined,linesnumbered,titlenotnumbered,noend]{algorithm2e}
\usepackage{paralist}
\usepackage{subcaption}

\usepackage{pgf}
\usepackage{tikz}

\usetikzlibrary{automata,positioning}
\usetikzlibrary{trees}
\usetikzlibrary{shapes}
\usetikzlibrary{petri}
\usetikzlibrary{arrows}
\usetikzlibrary{backgrounds}
\usetikzlibrary{calc}
\usetikzlibrary{fit}
\usetikzlibrary{decorations.pathmorphing}
\usetikzlibrary{decorations.text}
\usetikzlibrary{shapes.callouts}

\tikzstyle{rulenode}=[font=\footnotesize]
\tikzstyle{poedge}=[color=black,line width=1pt,->]
\tikzstyle{rfedge}=[color=blue,line width=1pt,->]
\tikzstyle{fredge}=[color=green,line width=1pt,->]
\tikzstyle{coedge}=[color=red,line width=1pt,->]
\tikzstyle{ponode}=[text=black]
\tikzstyle{rfnode}=[text=blue]
\tikzstyle{conode}=[text=red]
\tikzstyle{frnode}=[text=green]
\tikzstyle{enode}=[circle,fill,minimum size=5pt,inner sep=0pt]

\usepackage{colortbl}
\usepackage{multirow}
\usepackage{pifont}
\usepackage{tablefootnote}
\usepackage{listings}
\usepackage{cleveref}
\usetikzlibrary{backgrounds}
\usepackage[misc,geometry]{ifsym}

\author{Parosh Aziz Abdulla}
\affiliation{
  \institution{Uppsala University}            
   \country{Sweden}                    
}
\email{parosh@it.uu.se}          

\author{Mohamed Faouzi Atig}
\affiliation{
  \institution{Uppsala University}            
   \country{Sweden}                    
}
\email{mohamed_faouzi.atig@it.uu.se}         

\author{Bengt Jonsson}
\affiliation{
  \institution{Uppsala University}            
   \country{Sweden}                    
}
\email{bengt@it.uu.se}    

\author{Tuan Phong Ngo}
\affiliation{
  \institution{Uppsala University}            
   \country{Sweden}                    
}
\email{tuan-phong.ngo@it.uu.se} 

\begin{abstract}
We present a framework for the efficient application of stateless model checking
  (SMC) to concurrent programs running under the Release-Acquire (RA) fragment
  of the C/C++11 memory model. Our approach is based on exploring the possible
  {\it program orders}, which define the order in which instructions of a thread are executed, and
  {\it read-from} relations, which specify how reads obtain their values from writes.
  This is in contrast to previous approaches, which also explore the possible {\it coherence orders}, i.e., orderings between conflicting writes.
  Since unexpected test results such as program crashes or assertion violations
  depend only on the read-from relation, we avoid a
  potentially significant source of redundancy. Our framework is based on a novel
  technique for determining whether a particular read-from relation is
  feasible under the RA semantics. We define an SMC algorithm which is provably optimal
  in the sense that it explores each program order and read-from relation exactly once. This
  optimality result is strictly stronger than previous analogous optimality results,
  which also take coherence order into account. 
We have implemented our framework in
  the tool \textsc{Tracer}. Experiments show that \textsc{Tracer} can be significantly faster than
  state-of-the-art tools that can handle the RA semantics.
\end{abstract}

\ccsdesc[500]{Software and its engineering~Formal software verification}

\keywords{software model checking, weak memory models, C/C++11, Release-Acquire}  %% \keywords are mandatory in final camera-ready submission

\setcopyright{rightsretained}
\acmPrice{}
\acmDOI{10.1145/3276505}
\acmYear{2018}
\copyrightyear{2018}
\acmJournal{PACMPL}
\acmVolume{2}
\acmNumber{OOPSLA}
\acmArticle{135}
\acmMonth{11}

\begin{document}

\maketitle

\definecolor{mygreen}{rgb}{0.05, 0.5, 0.06}
\definecolor{myorange}{rgb}{0.93, 0.49, 0.1}
\definecolor{myred}{rgb}{0.82, 0.1, 0.26}
\definecolor{myblue}{rgb}{0.01, 0.28, 1.0}
\definecolor{myviolet}{rgb}{0.6, 0.4, 0.8}
\definecolor{mygray}{rgb}{0.9, 0.89, 0.89}
\definecolor{mypurple}{rgb}{0.41, 0.16, 0.38}

\newcommand\patodo[1]{\textcolor{red}{#1}}
\newcommand\phongtodo[1]{\textcolor{blue}{Phong: #1}}
\newcommand\faouzitodo[1]{\textcolor{orange}{#1}}
\newcommand\bjcom[1]{\textcolor{magenta}{#1}}
\newcommand\bjcomcom[1]{}

\newcommand\nat{{\mathbb N}}
\newcommand\alphabet\Sigma
\newcommand{\tuple}[1]{\left\langle#1\right\rangle}
\newcommand{\set}[1]{\left\{#1\right\}}
\newcommand{\setcomp}[2]{\left\{{#1}\mid {#2}\right\}}
\newcommand\aset{A}
\newcommand\bset{B}
\newcommand\fun{f}
\newcommand\sizeof[1]{|#1|}
\newcommand\app\bullet
\newcommand\wordsover[1]{{#1}^*}
\newcommand\word{w}
\newcommand\emptyword\epsilon
\newcommand\lengthof[1]{\mid\!\!#1\!\!\mid}
\newcommand\hdof[1]{{\it head}\!\left(#1\right)}
\newcommand\tailof[1]{{\it tail}\!\left(#1\right)}
\newcommand\lastof[1]{{\it last}\!\left(#1\right)}

\newcommand\hh{h}
\newcommand\ii{i}
\newcommand\jj{j}
\newcommand\kk{k}
\newcommand\mm{m}
\newcommand\nn{n}

\newcommand\aelem{a}
\newcommand\belem{b}
\newcommand\rel{R}
\newcommand\xvar{x}
\newcommand\yvar{y}
\newcommand\avar{{\tt a}}
\newcommand\bvar{{\tt b}}
\newcommand\cvar{{\tt c}}
\newcommand\dvar{{\tt d}}

\newcommand\init{{\tt init}}
\newcommand\initof[1]{\init_{#1}}
\newcommand\initx{\initof\xvar}

\newcommand\ra{{\rm RA}}
\newcommand\sra{{\tt SRA}}
\newcommand\scm{{\tt SC}}
\newcommand\nmodels\nvDash
\newcommand\restrict[2]{#1\raise-.5ex\hbox{\ensuremath|}_{#2}}

\newcommand\po{{\color{myred}{\tt po}}}
\newcommand\rf{{\color{mygreen}{\tt rf}}}
\newcommand\rfby[1]{\rf^{#1}}
\newcommand\rfx{\rfby\xvar}
\newcommand\rfy{\rfby\yvar}
\newcommand\urf{{\tt urf}}
\newcommand\urfby[1]{\urf^{#1}}
\newcommand\urfx{\urfby\xvar}
\newcommand\urfy{\urfby\yvar}
\newcommand\arwrf{{\tt RMWrf}}
\newcommand\arwrfby[1]{\arwrf^{#1}}
\newcommand\arwrfx{\arwrfby\xvar}
\newcommand\arwrfy{\arwrfby\yvar}
\newcommand\maxarwrf{{\tt RMWrf}_{\tt max}}
\newcommand\co{{\color{myorange}{\tt co}}}
\newcommand\coby[1]{\co^{#1}}
\newcommand\cox{\coby\xvar}
\newcommand\coy{\coby\yvar}
\newcommand\canco[1]{\widehat{{\co}(#1)}}
\newcommand\fr{{\color{myblue}{\tt fr}}}
\newcommand\frby[1]{\fr^{#1}}
\newcommand\frx{\frby\xvar}
\newcommand\fry{\frby\yvar}
\newcommand\acyclic{{\tt acyclic}}
\newcommand\eventsetof[1]{#1\!.\eventset}
\newcommand\poof[1]{#1.\po}
\newcommand\rfof[1]{#1.\rf}
\newcommand\coof[1]{#1.\co}
\newcommand\coxof[1]{#1.\cox}
\newcommand\extcoofx[1]{\langle{\cox}(#1)\rangle^+}

\newcommand\prog{{\mathcal P}}
\newcommand\memmodel{{\mathcal M}}
\newcommand\conf\gamma
\newcommand\terminatedconf{\conf_\downarrow}
\newcommand\runs{{\it Runs}}
\newcommand\truns{{\it TRuns}}
\newcommand\traces{{\tt Traces}}
\newcommand\istotal[1]{{\tt total}(#1)}
\newcommand{\idenotationof}[2]{\left\llbracket{#2}\right\rrbracket^{\tt I}_{#1}}
\newcommand{\ssdenotationof}[2]{\left\llbracket{#2}\right\rrbracket^{\tt Total}_{#1}}
\newcommand{\wdenotationof}[2]{\left\llbracket{#2}\right\rrbracket^{\tt Weak}_{#1}}
\newcommand{\pdenotationof}[2]{\left\llbracket{#2}\right\rrbracket^{\tt P}_{#1}}
\newcommand{\cdenotationof}[2]{\left\llbracket{#2}\right\rrbracket^{\tt S}_{#1}}
\newcommand{\ddenotationof}[2]{\left\llbracket{#2}\right\rrbracket^{\tt DPOR}_{#1}}
\newcommand\threadset{{\mathcal T}}
\newcommand\thread{{\it th}}
\newcommand\regsetof[1]{{\tt Reg}_{#1}}
\newcommand\reg{{\tt a}}
\newcommand\areg{{\tt a}}
\newcommand\breg{{\tt b}}
\newcommand\creg{{\tt c}}
\newcommand\regstate{{\mathcal R}}
\newcommand\thstate{\sigma}
\newcommand\run\rho
\newcommand\thrun\pi
\newcommand\pth\pi
\newcommand\varset{{\mathbb X}}
\newcommand\arwtype{{\tt RMW}}
\newcommand\utype{{\tt RMW}}
\newcommand\wtype{{\tt W}}
\newcommand\rtype{{\tt R}}
\newcommand\trace\tau
\newcommand\otrace\sigma
\newcommand\emptytrace{\trace_\emptyset}
\newcommand\emptyotrace{\otrace_\emptyset}
\newcommand\tlub\sqcup
\newcommand\tglb\sqcap
\newcommand\tequiv\sim
\newcommand\ttequiv\equiv
\newcommand\ctordering\sqSubset
\newcommand\tordering\sqsubseteq
\newcommand\stordering\sqsubset
\newcommand\eventset{{\tt E}}
\newcommand\initeventset{\eventset_{\it init}}
\newcommand\event{e}
\newcommand\newevent{{\it nEvent}}
\newcommand\neweventof[1]{\newevent\left(#1\right)}
\newcommand\swap{\texttt{Swappable}}
\newcommand\swapof[1]{\swap\!\left(#1\right)}
\newcommand\idof[1]{{#1}.{\it id}}
\newcommand\threadof[1]{{#1}.{\it thread}}
\newcommand\typeof[1]{{#1}.{\it type}}
\newcommand\varof[1]{{#1}.{\it var}}
\newcommand\valof[1]{{#1}.{\it val}}
\newcommand\srcof[1]{{#1}.{\it src}}
\newcommand\id{{\it id}}
\newcommand\type{\mathfrak{t}}
\newcommand\tracetuple{\tuple{\eventset,\po,\rf,\co}}
\newcommand\wtracetuple{\tuple{\eventset,\po,\rf}}
\newcommand\compose{;}

\newcommand\oneless[1]{#1\!-\!1}
\newcommand\zeroval{0}
\newcommand\valset{{\mathbb V}}
\newcommand\val{v}
\newcommand\sat{{\it sat}}
\newcommand\satof[1]{\sat\left({#1}\right)}
\newcommand\osatof[1]{\sat^1\left({#1}\right)}
\newcommand\tsatof[1]{\sat^2\left({#1}\right)}
\newcommand\varsatof[2]{{\it sat}_{#1}\left({#2}\right)}
\newcommand\xcosatof[1]{{\it cosat}_{\xvar}\left({#1}\right)}
\newcommand\cosatof[1]{{\it cosat}\left({#1}\right)}
\newcommand\xocosatof[1]{{\it cosat}^1_{\xvar}\left({#1}\right)}
\newcommand\ocosatof[1]{{\it cosat}^1\left({#1}\right)}
\newcommand\xtcosatof[1]{{\it cosat}^2_{\xvar}\left({#1}\right)}
\newcommand\tcosatof[1]{{\it cosat}^2\left({#1}\right)}
\newcommand\weak{{\it weak}}
\newcommand\weakof[1]{\weak\left(#1\right)}
\newcommand\refinedby{\sqsubseteq}
\newcommand\refinedeq{\equiv}
\renewcommand\ln{{\tt ln}}
\newcommand\lnof[1]{\ln\left(#1\right)}
\newcommand\plnof[2]{\ln^{#1}\left(#2\right)}
\newcommand\xlnof[1]{\plnof\xvar{#1}}
\newcommand\tlnof[1]{{\tt LN}\left(#1\right)}
\newcommand\tln{{\tt LN}}
\newcommand\ptlnof[2]{\tln^{#1}\left(#2\right)}
\newcommand\xtlnof[1]{\ptlnof\xvar{#1}}
\newcommand\tr{{\tt tr}}
\newcommand\trof[1]{\tr\left(#1\right)}
\newcommand\ptrof[2]{{\tt tr}^{#1}\left(#2\right)}
\newcommand\xtrof[1]{\ptrof\xvar{#1}}
\newcommand\ttr{{\tt TR}}
\newcommand\ttrof[1]{\ttr\left(#1\right)}
\newcommand\pttrof[2]{{\tt TR}^{#1}\left(#2\right)}
\newcommand\xttrof[1]{\pttrof\xvar{#1}}
\newcommand\add\odot

\newcommand\maxpoof[1]{{\tt MaxPO}\left(#1\right)}
\newcommand\visible{{\mathcal V}}
\newcommand\readable{{\mathcal R}}
\newcommand\avval{{\it AVal}}

\newcommand\extend\oplus
\newcommand\ssextend{\extend^{\tt SS}}
\newcommand\weakextend{\extend^{\tt W}}
\newcommand\wextend{\extend^{\wtype}}
\newcommand\rextend{\oplus^{\rtype}}
\newcommand\arwextend{\oplus^{\arwtype}}
\newcommand\uextend{\oplus^{\utype}}
\newcommand\extendof[1]{\extend\left(#1\right)}
\newcommand\lbl\ell
\newcommand\action\lbl
\newcommand\silentlbl\varepsilon
\newcommand\expr{e}
\newcommand\fvarsof[1]{{\tt FV}\left(#1\right)}
\newcommand\subst[3]{{#1}\{{#3}/{#2}\}}
\newcommand\cmdset{{\mathcal C}}
\newcommand\cmd{c}
\newcommand\skipcmd{{\tt skip}}
\newcommand\ifcmd{{\tt if}}
\newcommand\elsecmd{{\tt else}}
\newcommand\thencmd{{\tt then}}
\newcommand\repeatcmd{{\tt repeat}}
\newcommand\untilcmd{{\tt until}}
\newcommand\cascmd{{\tt CAS}}
\newcommand\fenceop{{\tt fence()}}
\newcommand\fvar{f}
\newcommand\valmap{\theta}
\newcommand\movesto[1]{\xrightarrow{#1}{}}
\newcommand\Movesto[1]{\xRightarrow{#1}{}}
\newcommand\imovesto[1]{\xrightarrow{#1}_{\tt I}}
\newcommand\iMovesto[1]{\xRightarrow{#1}_{\tt I}}
\newcommand\ssmovesto[1]{\xrightarrow{#1}_{\tt SS}}
\newcommand\ssMovesto[1]{\xRightarrow{#1}_{\tt SS}}
\newcommand\pmovesto[1]{\xrightarrow{#1}_{\tt P}}
\newcommand\pMovesto[1]{\xRightarrow{#1}_{\tt P}}
\newcommand\cmovesto[1]{\xrightarrow{#1}_{\tt S}}
\newcommand\succof[1]{{\tt succ}\!\left(#1\right)}
\newcommand\csuccof[1]{{\tt succ}_{\tt S}\!\left(#1\right)}
\newcommand\cMovesto[1]{\xRightarrow{#1}_{\tt S}}
\newcommand\lmovesto[1]{\stackrel{#1}\leadsto}
\newcommand\assigned\leftarrow
\newcommand\freshof[1]{{\it fresh}\left(#1\right)}

\newcommand\rschedule{\textsc{RunSchedule}}
\newcommand\explore{\textsc{DfVisit}}
\newcommand\declarepostponed{\textsc{DeclarePostponed}}
\newcommand\visited{V}
\newcommand\scheduledof[1]{\texttt{Schedules}\!\left(#1\right)}
\newcommand\myposition{\texttt{RMW-VisitPosition}}
\newcommand\positionof[1]{\myposition\!\left(#1\right)}
\newcommand\eorder{\alpha}
\newcommand\orderingof[1]{\ordering\left(#1\right)}
\newcommand\result{result}
\newcommand\schedule\beta
\newcommand\nevent{{\it NewE}}
\newcommand\neventof[1]{\nevent\left(#1\right)}

\newcommand\ecnt{{\tt ecnt}}

\newcommand\traceof[1]{{\it trace}\left(#1\right)}
\newcommand\confof[1]{{\it conf}\left(#1\right)}
\newcommand\orderof[1]{{\it order}\left(#1\right)}

\newcommand\setname[1]{{\mathcal A}}

\newcommand\inity{\initof\yvar}

\newcommand\true{{\tt true}}
\newcommand\false{{\tt false}}

\newcommand\preof[1]{{\it Pre}\left(#1\right)}
\newcommand\penables{\vdash_{\tt P}}

\newcommand\cpenables{\mathrel{{\tikz[anchor=base,baseline]{\node[name=vdash]{$\vdash_{\tt P}$};\node[name=cdot,anchor=center] at ($(vdash.center)+(-2pt,2pt)$) {$\cdot$};}}}}
\newcommand\cdenables{\tikz[anchor=base,baseline]{\node[name=vdash]{$\vdash_{\tt D}$};\node[name=cdot,anchor=center,circle,fill=black,inner sep=0pt,minimum size=2pt] at ($(vdash.center)+(-1.7pt,2pt)$) {};
}}
\newcommand\denables{\vdash_{\tt D}}
\newcommand\obs\alpha
\newcommand\obsseq\pi
\newcommand\obsseqsub\preceq
\newcommand\obsseqminus\ominus
\newcommand\obsdep{\lhd}
\newcommand\obsequiv{\sim}
\newcommand\obsseqequiv{\approx}
\newcommand\obsseqordering{\sqSubset}
\newcommand\eventof[1]{#1.{\it event}}

\newcommand\existsfunc{\texttt{exists}}
\newcommand\existsfuncof[2]{\existsfunc(#1,#2)}

\crefformat{section}{\S#2#1#3} 
\crefformat{subsection}{\S#2#1#3}
\crefformat{subsubsection}{\S#2#1#3}
\crefformat{subsubsection}{\S#2#1#3}
\crefformat{appendix}{Appendix~#2#1#3}
\crefformat{subappendix}{Appendix~#2#1#3}
\crefformat{subsubappendix}{Appendix~#2#1#3}

\definecolor{mGreen}{rgb}{0,0.6,0}
\definecolor{mGray}{rgb}{0.5,0.5,0.5}
\definecolor{mPurple}{rgb}{0.58,0,0.82}
\definecolor{backgroundColour}{rgb}{0.95,0.95,0.92}

\lstdefinestyle{CStyle}{
    backgroundcolor=\color{white},   
    commentstyle=\color{mGreen},
    keywordstyle=\color{magenta},
    numberstyle=\tiny\color{mGray},
    stringstyle=\color{mPurple},
    basicstyle=\linespread{0.8}\small,
    breakatwhitespace=false,         
    breaklines=true,                 
    captionpos=b,                    
    keepspaces=true,                 
    numbers=left,                    
    numbersep=5pt,                  
    showspaces=false,                
    showstringspaces=false,
    showtabs=false,                  
    tabsize=2,
    language=C,
    frame = single,
    framexleftmargin=15pt
}

\SetAlgoNoLine
\SetInd{0.8em}{0.8em}
\SetNoFillComment
\newcommand\mycommfont[1]{\footnotesize\ttfamily\textcolor{mGreen}{#1}}
\SetCommentSty{mycommfont}
\DontPrintSemicolon

\newcommand\setnamefive{{\mathcal R}\texttt{-ForRead}}
\newcommand\setnamefour{{\mathcal R}\texttt{-ForRMW}}
\newcommand\setnameone{\texttt{Max-ReadFroms}}
\newcommand\setnametwo{\texttt{Max-Sources}}
\newcommand\setnamethree{\visible}
\newcommand\tnode{n}

\section{Introduction}
\label{sect:introduction}

Ensuring correctness of concurrent programs is difficult
since one must consider all the different ways in which
threads can interact.
A successful technique for finding concurrency bugs (i.e., defects that
arise only under some thread schedulings), and for
verifying their absence, is
\emph{stateless model checking} (SMC)~\cite{Godefroid:popl97}.
Given a terminating program, which may be annotated with assertions,
SMC systematically explores the set of
all thread schedulings that are possible during runs of this program.
A special runtime scheduler drives the SMC exploration by making decisions
on scheduling whenever such choices may affect the interaction
between threads; so that the exploration covers all possible
executions and detects any unexpected program results, program crashes,
or assertion violations. The technique is entirely automatic, has no
false positives, does not consume excessive memory, and can quickly
reproduce the concurrency bugs it detects.
SMC has been
implemented in tools, such as VeriSoft~\cite{Godefroid:verisoft-journal},
\textsc{CDSChecker}~\cite{DBLP:conf/oopsla/DemskyL15,NoDe:toplas16},
\textsc{Chess}~\cite{MQBBNN:chess}, Concuerror~\cite{Concuerror:ICST13},
rInspect~\cite{DBLP:conf/pldi/ZhangKW15}, and Nidhugg~\cite{tacas15:tso}, and
successfully applied to realistic concurrent
programs~\cite{GoHaJa:heartbeat,KoSa:spin17}.

SMC faces the problem that
the number of possible thread schedulings grows exponentially with
the length of program execution, and
must therefore be equipped with techniques to reduce the number of
explored executions.
The most prominent one is
\emph{partial order reduction}~\cite{Valmari:reduced:state-space,Peled:representatives,Godefroid:thesis,CGMP:partialorder},
adapted to SMC as \emph{dynamic partial order reduction} (DPOR).
DPOR was first developed for
concurrent programs that execute under the standard model of
Sequential Consistency (SC)~\cite{FG:dpor,SeAg:haifa06,abdulla2014optimal}. In recent years, DPOR has been adapted to
hardware-induced weak memory models, such as
TSO and PSO~\cite{tacas15:tso,DBLP:conf/pldi/ZhangKW15}, and language-level
concurrency models, such as the C/C++11 memory model~\cite{NoDe:toplas16,KLSV:popl18}.
DPOR is based on the observation that 
two executions can be regarded as equivalent if they induce the same ordering
between conflicting statement executions (called events),
and that it is therefore sufficient to
explore at least one execution in each equivalence class.
Under SC, such equivalence classes are called \emph{Mazurkiewicz~traces}~\cite{Mazurkiewicz:traces}; for weak memory models, the natural generalization
of Mazurkiewicz traces are called \emph{Shasha-Snir traces}~\cite{ShSn:parallel}.
A Shasha-Snir trace characterizes an execution of a
program by three relations between events;
\begin{inparaenum}[(i)]
\item
  $\po$ (``program order'') totally orders the events of each thread,
\item
    $\co$ (``coherence'') totally orders the writes to each shared variable, and
\item
    $\rf$ (``read-from'') connects each write with the reads that read its value.
\end{inparaenum}
Under weak memory models, the $\co$ and $\rf$ relations need not be
derived from the global order in which events occur in an execution
(as is the case under SC).
Each particular weak memory model therefore imposes restrictions on how
these relations may be combined.

\begin{figure}[tb]
 \begin{minipage}[c]{0.25\linewidth}

  \subcaptionbox{\label{overview:prog:fig}}{%
         
      \begin{tikzpicture}[line width=1pt,framed]
	\node[name=n0] at (0,0) {Initially: $\xvar=0$};

	\node[name=n11,anchor=west] at ($(n0.east)+(-75pt,-35pt)$) {$\xvar:=1$};
	\node[name=n12,anchor=north west] at ($(n11.south west)+(0pt,-5pt)$) {$\avar:=\xvar$};

	\node[name=n21,anchor=west] at ($(n11.east)+(15pt,0pt)$) {$\xvar:=2$};
	\node[name=n22,anchor=north west] at ($(n21.south west)+(0pt,-4pt)$) {$\bvar:=\xvar$};

	\node[name=th1,anchor=south west]at ($(n11.north west)+(5pt,1pt)$) {$\thread_1$};
	\node[name=th2,anchor=south west]at ($(n21.north west)+(5pt,1pt)$) {$\thread_2$};

	\draw[line width= 0.5pt] ($(n11.north east)+(7.5pt,10pt)$) -- ($(n11.south east)+(7.5pt,-20pt)$);
	\draw[line width= 0.5pt] ($(n11.north east)+(10.5pt,10pt)$) -- ($(n11.south east)+(10.5pt,-20pt)$);

	%\node[fit=(th1),outer sep=0pt,inner sep = -0.3pt, line width=0.5pt,draw] {};
	%\node[fit=(th2),outer sep=0pt,inner sep = -0.3pt,line width=0.5pt,draw] {};

	\end{tikzpicture}
  }
  \end{minipage}
  \hspace{5mm}
  \begin{minipage}[c]{0.6\linewidth}
  \center
  \subcaptionbox{\label{overview:SS:prog:fig}}{%
      \includegraphics{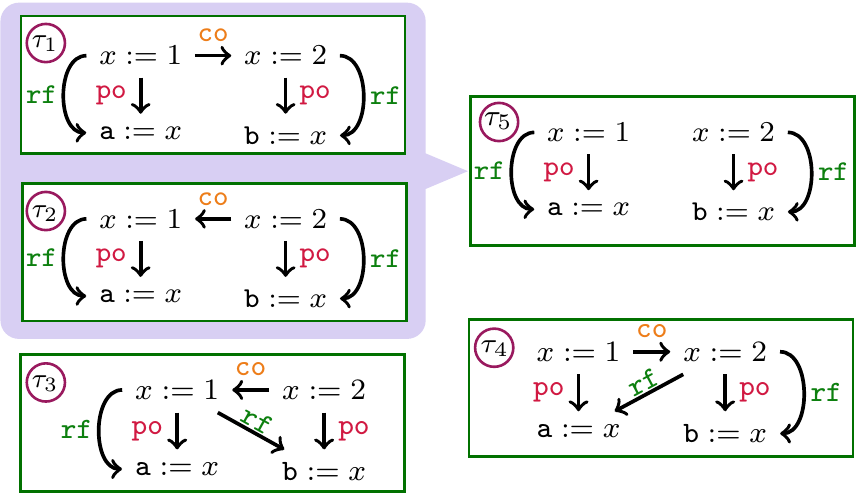}
  }
  \end{minipage}
%  \vspace{-2mm}
  \caption{ (a) A simple concurrent program and (b) 
    Shasha-Snir traces and weak traces.}
\end{figure}
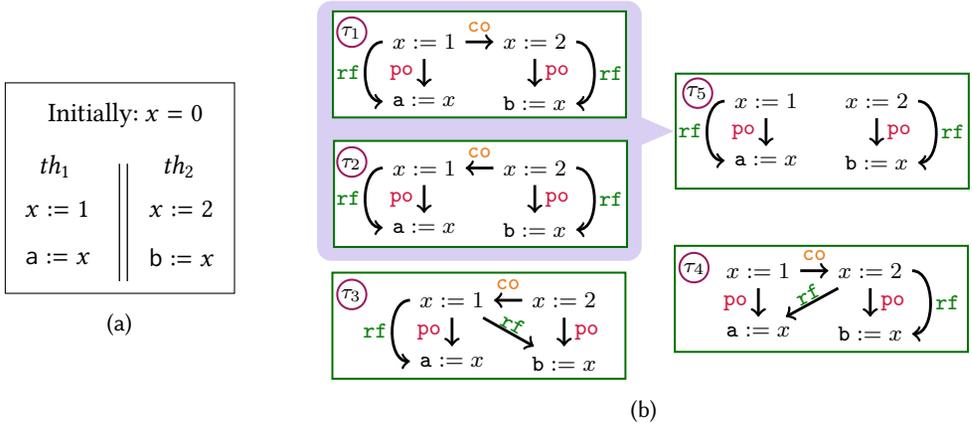

As an illustration,
Figure~\ref{overview:prog:fig} shows a simple program with two threads,
$\thread_1$ and $\thread_2$, that communicate through a shared variable $\xvar$.
Each thread writes to the  variable and reads from it into a local register, $\avar$ resp.\ $\bvar$.
We would like to explore the possible executions of this program, e.g., to
check whether the program can
satisfy $\avar=2$ and $\bvar=1$ upon termination.
Under many memory models, including SC, TSO, PSO, RA, and POWER,
executions of the program in Figure~\ref{overview:prog:fig} fall into four
equivalence classes, represented by the four possible Shasha-Snir traces
$\trace_1$, $\trace_2$, $\trace_3$, and $\trace_4$
in Figure~\ref{overview:SS:prog:fig}\footnote{A Shasha-Snir trace also includes events
that write initial values, but these can be ignored for this example.}.
A DPOR algorithm based on Shasha-Snir traces
(e.g.,~\cite{abdulla2014optimal,DBLP:conf/cav/AbdullaAJL16})
must thus explore at least four executions.
However, it is possible to reduce this number further.
Namely, a closer inspection reveals that the two traces
$\trace_1$ and $\trace_2$ are equivalent,
in the sense that each thread goes through the same sequences of local states
and computes the same results. This is because $\trace_1$ and $\trace_2$
have the same program order ($\po$) and read-from ($\rf$) relation.
Their only difference is how writes are ordered by $\co$, but this is
not relevant for the computed results.

The preceding example illustrates that
there is a potential for improving the efficiency of DPOR
algorithms by using a weaker equivalence induced only by $\po$ and
$\rf$. In this example, the improvement is modest (reducing the number of
explored traces from four to three), but it can be significant,
sometimes even exponential, for more extensive programs.
Several recent DPOR techniques try to exploit the potential offered by
such a weaker equivalence~\cite{NoDe:toplas16,DBLP:conf/pldi/Huang15,DBLP:conf/oopsla/Huang016,DC-DPOR@POPL-18}.
However, except for the minimal case of an acyclic communication graph~\cite{DC-DPOR@POPL-18},
they are far from optimally doing this, since 
they may still explore a significant number of different executions
with the same $\rf$ relation.
Therefore, the challenge remains to define a more efficient DPOR algorithm, 
that is optimal in the sense that it explores precisely one execution in
each equivalence class induced by $\po$ and $\rf$.
%% One of the difficulties that arises is to only explore those combinations
%% of $\po$ and $\rf$ that can arise in an execution;
%% %% for instance, in the program of
%% %% Fig.~\ref{overview:prog:fig} most relaxed memory models forbid
%% e.g., for the program in Fig.~\ref{overview:prog:fig}, letting
%% both threads read from the write of the other thread is forbidden under most
%% relaxed memory models.

In this paper, we present a fundamentally new approach to defining
DPOR algorithms, which optimally explores only the equivalence classes
defined by the program order and read-from relations.
Our method is developed for the Release-Acquire
(RA) fragment \cite{DBLP:conf/popl/LahavGV16} of the C/C++11 memory model.
RA  is a useful and well-behaved fragment of the C/C++11 memory model, which
strikes a good balance between performance and programmability.
In the RA semantics, all writes are \texttt{release} accesses, while all reads are \texttt{acquire} accesses.
RA allows high-performance  implementations, while still providing sufficiently
strong guarantees for fundamental concurrent algorithms
(such as the read-copy-update mechanism~\cite{DBLP:conf/popl/LahavGV16}).
Our DPOR algorithm is based on the above weakening of  Shasha-Snir traces,
called {\em weak traces}, which
are defined by only the program order and read-from relations of an execution.
For example, the program in Figure~\ref{overview:prog:fig} has three weak
traces, shown as $\trace_3$, $\trace_4$, and $\trace_5$.
Our DPOR algorithm is provably optimal for weak traces, in
the sense that it explores {\em precisely} one execution for each weak trace
that is 
{\em RA-consistent}, i.e., that can be extended with some coherence relation
that satisfies the constraints of the RA semantics.

A significant challenge for our DPOR algorithm is to efficiently
determine those continuations of a currently explored trace that lead to
some RA-consistent trace.
E.g., for the program in Figure~\ref{overview:prog:fig}, letting
both threads read from the write of the other thread leads to RA-inconsistency.
We solve this problem by defining
a {\em saturation} operation, which extends a weak trace with a \emph{partial}
coherence relation, which contains precisely
those coherence edges that must be present in any
corresponding Shasha-Snir trace.
During exploration, the DPOR algorithm
maintains a saturated version of the currently explored
weak trace, and can therefore examine
{\em precisely} those weak traces that are RA-consistent,
without performing useless explorations.
When a read event is added, the algorithm determines the set of
write events from which it can obtain its value while preserving
RA-consistency, and branches into a separate continuation for each
such write event.
When a write event is added, the algorithm merely adds it to the
trace. It can be proven that this preserves RA-consistency, and also keeps the
trace saturated (a slight modification is needed for atomic read-modify-write
events). The algorithm must also detect if some previous read may
read from a newly added write, and then backtrack  to allow the write
to be performed before that read.

We  prove that our DPOR algorithm does not perform
any useless work, in the sense that
\begin{inparaenum}[(i)]
\item
  any exploration eventually leads to a terminated
  RA-consistent execution, i.e., the algorithm never blocks because it discovers
  that it is about to perform redundant or wasted explorations, and
\item
  each RA-consistent weak trace is explored precisely once.
\end{inparaenum}
%In the appendix (space does not permit inclusion in the main paper),
%\phongtodo{How to rewrite this sentence?}
%we present an extension of the saturation technique
%and of the DPOR algorithm to handle
%atomic read-modify-write (RMW) operations and locks. 
Our saturation technique
and the DPOR algorithm 
presented in this paper
can be extended to cover 
atomic read-modify-write (RMW) operations and locks
(space does not permit inclusion in this paper).
This extension also
satisfies the same strong optimality results (in (i) and (ii)).

In summary, we prove a stronger notion of optimality than related SMC
approaches: we prove optimality concerning weak traces, whereas others
only prove it w.r.t.\ total traces (i.e.,\ Mazurkiewicz traces or Shasha-Snir traces);
moreover,
our technique can be extended to cover RMWs. 
%we are the first to cover
%RMW operations in such a result.

%We demonstrate the effectiveness of our approach by implementing \textsc{Tracer}, a
%static model checker for C programs running under RA.
%Our experiments confirm that \textsc{Tracer} outperforms comparable tools on benchmarks
%%with a significantly small number of weak traces than total traces.
%%On benchmarks where these numbers are comparable, also our tool has
%with a significantly smaller number of weak traces than total traces.
%On benchmarks where these numbers are comparable, also \textsc{Tracer} has
%comparable performance.

We have implemented our saturation operation
and
DPOR algorithm
in a tool, called \textsc{Tracer},
and applied
it to many challenging  benchmarks.
We compare our tool
with other state-of-the-art stateless model checking tools
running under the RA semantics.
The experiments
show that
\textsc{Tracer}
always
generates 
optimal numbers of executions w.r.t.\ weak traces in all benchmarks.
On  many benchmarks, this number is much smaller than
the ones produced by the other tools.
The results
also show that
\textsc{Tracer}
has better performance
and scales better
 for more extensive programs,
even in the case where it 
explores the same number of executions
as the other tools.

%%
%\paragraph{Outline}
%%\patodo{To be updated.}
%%The next section illustrates our main concepts on a simple example.
%In Section~\ref{prels:section}, we introduce some notation.
%In Section~\ref{model:section}, we describe concurrent programs, the RA semantics, weak traces, and 
%the semantics based on weak traces.
%%
%In
%Section~\ref{saturated:section}, we 
%describe how traces are saturated during exploration.
%%
%We describe  our DPOR algorithm in  
%Section~\ref{dpor:section}.
%%
%%In \cref{atomic:section}, we cover ARW operations.
%%
%We present our  experimental results in Section~\ref{experiment:section}.
%%
%Finally we describe related work and conclusions in Section~\ref{related:section} and Section~\ref{conclusions:section} respectively.
%%
%%In the supplementary material, 
%%we provide an appendix that contains proofs of all the lemmas, a description of running our DPOR algorithm on an example, and details about the benchmarks used in our experimentation.
%%%}

\section{Basic Notation}
\label{prels:section}

We let $\nat$ denote the set of natural numbers.
Fix a set $\aset$.
If $\aset$ is finite then we use $\sizeof\aset$ to denote the
size of $\aset$.
For a binary relation $\rel$ on $\aset$,
we write $\aelem_1\;[\rel]\;\aelem_2$ to denote that
$\tuple{\aelem_1,\aelem_2}\in\rel$.
We use $\rel^{-1}$
to denote the inverse of $\rel$, i.e. 
$\aelem_1\;[\rel^{-1}]\;\aelem_2$ iff
$\aelem_2\;[\rel]\;\aelem_1$.
We use $\rel^+$ and $\rel^*$ to denote the transitive closure 
and the reflexive transitive closure of $\rel$, respectively.
We 
write $\aelem_1\;[\rel]^+\aelem_2$ and $\aelem_1\;[\rel]^*\aelem_2$ 
to denote that
$\aelem_1\;[\rel^+]\;\aelem_2$
resp.\ $\aelem_1\;[\rel^*]\;\aelem_2$.
We say that $\rel$ is a partial order if it is irreflexive
(i.e., $\neg(\aelem\;[\rel]\;\aelem)$ for all $\aelem\in\aset$)
 and transitive
(i.e., if $\aelem_1\;[\rel]\;\aelem_2$  and
$\aelem_2\;[\rel]\;\aelem_3$ then
$\aelem_1\;[\rel]\;\aelem_3$
for all $\aelem_1,\aelem_2,\aelem_3\in\aset$).
We say that $\rel$ is total if, for all 
$\aelem_1,\aelem_2\in\aset$, either
$\aelem_1\;[\rel]\;\aelem_2$ or
$\aelem_2\;[\rel]\;\aelem_1$.
We use $\acyclic(\rel)$ to denote that
$\rel$ is acyclic,
i.e., there is no $\aelem\in\aset$
such that 
$\aelem\;[\rel]^+\;\aelem$.
For a set $\bset\subseteq\aset$, we define 
$\restrict\rel\bset:=\rel\cap(\bset\times\bset)$, i.e.,
it is the restriction  of $\rel$ to $\bset$.
For binary relations $\rel_1$ and $\rel_2$ on $\aset$, we use
$\rel_1\compose\rel_2$ to denote the composition of $\rel_1$ and
$\rel_2$, i.e., $\aelem_1\;[\rel_1\compose\rel_2]\;\aelem_2$
iff there is an $\aelem_3\in\aset$ such that
$\aelem_1\;[\rel_1]\;\aelem_3$ and
$\aelem_3\;[\rel_2]\;\aelem_2$.
For sets $\aset$ and $\bset$, we use
$\fun:\aset\to\bset$ to denote that $\fun$ is a 
(possibly partial) function
from $\aset$ to $\bset$.
%
%We write $\fun(\aelem)=\undf$ to denote that the value of $\fun$ is undefinedfor $\aelem$.
%
We use $\fun[\aelem\leftarrow\aelem']$ to denote the function
$\fun'$ such that $\fun'(\aelem)=\aelem'$ and
$\fun'(\belem)=\fun(\belem)$ if $\belem\neq\aelem$.
We use $\wordsover\aset$ to denote the set of finite words over $\aset$,
and use $\emptyword$ to denote the empty word.
We use $\lengthof\word$ to denote the length of $\word$,
use $\word[\ii]$ to denote the $i^{th}$ element of $\word$, and
use $\lastof\word$ for $\word[\lengthof\word]$.
%
%% We define $\hdof\word:=\word[1]$, 
%% $\tailof\word:=\word[2]\cdots\word[\lengthof\word]$.
%
For words $\word_1,\word_2\in\wordsover\aset$, we use
$\word_1\app\word_2$ to denote the concatenation
$\word_1$ and $\word_2$.

\section{Model}
\label{model:section}

\paragraph{Programs}
We consider a program $\prog$
consisting of a finite set $\threadset$  of {\em threads} 
({\it processes}) that share 
a finite set $\varset$ of {\em (shared) variables}, ranging over 
a domain $\valset$ of {\em values} that includes a special
value $\zeroval$.
A thread  has a finite set
of local registers that store values from $\valset$.
Each thread runs a deterministic code, built in a standard way
from expressions and atomic commands, using standard control
flow constructs (sequential composition, selection, and bounded loop constructs).
Throughout the paper, we use $\xvar, \yvar$ for shared variables,
$\areg, \breg, \creg$ for registers, and $\expr$ for expressions.
{\it Global statements}
are either  write $\xvar:=\expr$ to a shared variable or
read $\reg:=\xvar$ from a shared variable.
{\it Local statements} only access and affect the local state
of the thread and include 
assignments $\reg:=\expr$ to registers
and conditional control flow constructs.
Note that expressions do not contain shared variables, implying that a statement
accesses at most one shared variable.
%\bjcom{Is this better?}
%\patodo{Perhaps}
%
For readability reason, 
we do not consider 
atomic read-modify-write (RMW) operations and fences.
%In Section~\ref{atomic:section}, we  consider
%atomic read-modify-write (RMW) operations and fences.
%
%\phongtodo{We need to fix this sentence because of removing appendix}
%
The local state of a thread $\thread\in\threadset$ 
is defined as usual by its program counter and the contents
of its registers.

\paragraph{Configurations}
A {\it configuration (global state)} of $\prog$
is made up of the local states of all the threads.
Note that the values of the shared variables
are not part of a configuration.
The reason is that, under the RA semantics (and many other weak memory models), 
different threads may have different ``local views'' of the shared variables,
i.e., they may see different values of the shared variables 
at a given point during the program execution.
Therefore, it is not possible to assign a unique value to a shared variable.
In existing operational semantics for such weak memory models
including the RA semantics
(e.g.,~\cite{DBLP:conf/ecoop/KaiserDDLV17}),
a write instruction does not explicitly modify the
values of the shared variables. 
Instead, a write instruction is added to a ``pool'' of write instructions
that have been issued by the threads.
A read instruction can fetch its value from a set of available
write instructions in the pool and update its local view accordingly.
The sets of write instructions that are available
to reading threads depend on the particular memory model.

Following~\citet{DBLP:conf/popl/LahavGV16},
we define an operational semantics for $\prog$
as a labeled transition relation over configurations.
%
%% \patodo{I have added ``following the taming paper here''}
%
Each transition corresponds to one thread performing a local or global statement.
A transition between two configurations $\conf$ and $\conf'$ is of form
$\conf\lmovesto{\lbl}\conf'$,
where the label $\lbl$ describes the interaction with shared variables.
The label $\lbl$ is one of three forms:
\begin{inparaenum}[(i)]
(i)
$\tuple{\thread,\silentlbl}$,
  indicating  a local statement performed by thread $\thread$, which
  updates only the local state of $\thread$,
%% the transition updates the program counter and the registers of $\thread$,
(ii)
$\tuple{\thread,\wtype,\xvar,\val}$,
indicating a write of the value $\val$ to the variable $\xvar$
by the thread $\thread$, which also
updates the program counter of $\thread$,
and
(iii)
$\tuple{\thread,\rtype,\xvar,\val}$ 
indicating a read of $\val$ from $\xvar$
by the thread $\thread$
into some register, while also updating the program counter of $\thread$.
\end{inparaenum}
Observe that since the shared variables
are not part of a configuration, 
the threads do not interact 
with each other in the transition relation.
In particular,
%% write statements do not affect the shared variables, and 
%% a read statement can read an arbitrary value.
there is no constraint on the values that
are used in transitions corresponding to read statements.
This will obviously make illegal program behaviors possible.
We remedy this problem by associating runs with so-called {\em traces},
which (among other things) represent how reads obtain their values from
writes. 
A particular memory semantics (such as the RA semantics)
is formulated by imposing 
restrictions on these traces, thereby also restricting the possible
runs that are associated with them.
%

%Also, note that, given the configuration $\conf$ and the thread $\thread$, the label $\lbl$ and the configuration $\conf'$ are unique due to the assumption that the code of $\thread$ is deterministic.

Since local statements are not visible to other
threads, we will not represent them explicitly 
in the transition relation  considered in our DPOR algorithm. Instead,
we let each transition represent the combined effect of some
finite sequence of local statements  by a thread followed
by a global statement by the same thread.
More precisely, for configurations $\conf$ and $\conf'$ and a label
$\lbl$ which is either of the form
$\tuple{\thread,\wtype,\xvar,\val}$ or of the form
$\tuple{\thread,\rtype,\xvar,\val}$, we let 
$\conf\movesto\lbl\conf'$ denote that
we can reach $\conf'$ from $\conf$ by performing 
a sequence of transitions labeled
with $\tuple{\thread,\silentlbl}$ followed by a transition labeled with $\lbl$.
Defining the  relation $\movesto{}$ in this manner 
 ensures that we take the effect of local
statements into account while
avoiding consideration of interleavings of
local statements of different threads in the analysis. 
Such 
optimization is common in tools (e.g., Verisoft~\cite{Godefroid:popl97}).
%\patodo{A thread may need to perform local transitions 
 % after the very last global statement in order reach a terminal
 % configuration.}

We introduce some extra notation.
We use $\conf\movesto{\tuple{\thread,\rtype,\xvar,*}}*$ 
to denote that
$\conf\movesto{\tuple{\thread,\rtype,\xvar,\val}}\conf'$
for some value $\val\in\valset$ and configuration $\conf'$.
We use $\conf\movesto{}\conf'$ to denote that
$\conf\movesto\lbl\conf'$ for some $\lbl$ and define
$\succof\conf:=
\setcomp{\conf'}
{\conf\movesto{}\conf'}$,
i.e., it is the set of successors of 
$\conf$ w.r.t.\  $\movesto{}\!$.

A configuration $\conf$ is said to be {\it terminal} if 
$\succof\conf=\emptyset$, i.e., no thread can execute a global statement from
$\conf$.
A {\it run} $\run$ from $\conf$ is a sequence
$\conf_0\movesto{\lbl_1}\conf_1
\movesto{\lbl_2}\cdots\movesto{\lbl_\nn}
\conf_\nn$ 
such that $\conf_0=\conf$.
We say that $\run$ is {\it terminated} if $\conf_\nn$ is terminal.
We let $\runs(\conf)$ denote the set of runs
from $\conf$.

\paragraph{Events}
An event corresponds to a particular execution
of a statement in a run of $\prog$.
A {\it write event} $\event$ is a tuple
$\tuple{\id,\thread,\wtype,\xvar,\val}$,
where 
$\id\in\nat$ is an event identifier, 
$\thread\in\threadset$ is a thread,
$\xvar\in\varset$ is a variable, and
$\val\in\valset$ is a value.
This event 
corresponds to thread $\thread$
writing the value $\val$ to variable $\xvar$.
The identifier $\id$ denotes that $\thread$ has
executed $\oneless{id}$ events before $\event$
%% fines the order in which $\event$ is executed among all
in the corresponding run.
A {\it read event} $\event$ is a tuple
$\tuple{\id,\thread,\rtype,\xvar}$,
where $\id$, $\thread$, and $\xvar$ are as for
a write event.
This event 
corresponds to thread $\thread$
reading some value to $\xvar$.
Note that a read event $\event$ does not specify the particular 
value it reads.
This value will be defined in a trace by specifying a write event 
from which $\event$ fetches its value.
%%% read by $\event$ is then equal to the value written by $\event'$.)
%
For an event $\event$ of form $\tuple{\id,\thread,\wtype,\xvar,\val}$ or
$\tuple{\id,\thread,\rtype,\xvar}$, we define 
$\idof\event:=\id$,
$\threadof\event:=\thread$,
$\typeof\event:=\type$ where $\type\in\set{\wtype,\rtype}$,
$\varof\event:=\xvar$, and
$\valof\event:=\val$ (the latter is not defined for a read event).
For each variable $\xvar\in\varset$, we assume
a special write event $\initx=\tuple{-,-,\wtype,\xvar,\zeroval}$,
called the {\it initializer} event for $\xvar$.
This event is not performed by any of the threads in $\threadset$, and
writes the value $\zeroval$ to $\xvar$.
We define $\initeventset:=\setcomp{\initx}{\xvar\in\varset}$ as
the set of initializer events.

If $\eventset$ is a set of events, we define
subsets of $\eventset$ characterized
by particular attributes of its events.
For instance, for a  thread $\thread$, we let $\eventset^{\thread,\wtype}$
denote $\setcomp{\event \in \eventset}
{\threadof\event=\thread \wedge \typeof\event=\wtype}$.

\paragraph{Traces}
A {\it trace}     
$\trace$ is a tuple $\tracetuple$,
where $\eventset$ is a set of {\it events}
which includes the set $\initeventset$ of initializer events,
and where $\po$ (program order),
$\rf$ (read-from), and
$\co$ (coherence order) are binary relations on $\eventset$ that satisfy:
\begin{itemize}

\item[{$\event\;[\po]\;\event'$}] if $\threadof{\event} = \threadof{\event'}$ and
  $\idof{\event} < \idof{\event'}$, i.e., $\po$
  totally orders the events of each
  individual thread.
  As mentioned above, each event corresponds to an execution
  of a program  statement.
  The program order then reflects the order in which the statements
  of a given thread are executed.
  Note that $\po$ does not relate the initializer events.
\item[{$\event\;[\rf]\;\event'$}] if $\event$ is a write event and
  $\event'$ is a read event on the same variable, which obtains its value from
  $\event$. The inverse of relation $\rf$, denoted
  $\rf^{-1}$, must be a total function on $\eventset^\rtype$.
  We sometimes view $\rf$ as the union of a read-from relation
  $\rfx$ for each variable $\xvar\in\varset$.
\item[$\co$] is a union
    $\co=\cup_{\xvar\in\varset}\cox$, 
  where $\cox$ is a relation on $\eventset^{\wtype,\xvar}$ (including
  $\initx$),  subject to the
  constraint that $\initx$   is before all other write events in
  $\eventset^{\wtype,\xvar}$.
  Thus $\co$ does not relate write events on different variables.
  The relation $\cox$ reflects how the threads view the order on the write 
  events  on $\xvar$.
  If $\event_1\;[\cox]\;\event_2$ then all threads share 
  the view that $\event_1$ has occurred before $\event_2$.  
\end{itemize}
Note that (in contrast to $\po$), each $\cox$ can be an arbitrary relation;
it need not even be transitive.
We say that a trace $\tracetuple$ is {\em total} if 
$\cox$ is a strict total order on $\eventset^{\wtype,\xvar}$ for each $\xvar\in\varset$.
We use $\istotal{\trace}$ to denote that $\trace$ is total.
A total trace is also called a {\it Shasha-Snir trace}.
We sometimes use {\em partial trace} to denote an arbitrary trace,
when we want to emphasize that it need not be total.

As depicted in Figure~\ref{overview:SS:prog:fig},
we can view $\trace = \tracetuple$ as a graph whose nodes
are $\eventset$ and whose edges are defined by the 
relations $\po$, $\rf,$ and $\co$. 
%We use the notation $\eventsetof{\trace}$ for the set of events of $\trace$, and analogously for $\poof{\trace}$, etc.
We define $\sizeof\trace:=\sizeof\eventset$, i.e., it is the
number of events in $\trace$.
We define the {\it empty trace}
$\emptytrace:=\tuple{\initeventset,\emptyset,\emptyset,\emptyset}$,
i.e., it contains only the initializer events,
and all the relations are empty.

\paragraph{Associating Traces with Runs}
We can now define when a trace can be associated with a run.
Consider a run 
$\run$ of form 
$\conf_0\movesto{\lbl_1}\cdots\movesto{\lbl_\nn}\conf_\nn$,
where $\lbl_\ii=\tuple{\thread_\ii,\type_\ii,\xvar_\ii,\val_\ii}$,
and let $\trace=\tracetuple$ be a trace.
We write $\run\models\trace$ to denote that
the following conditions are satisfied:

%\medskip
%\noindent
%$\bullet$ 
\begin{itemize}
\item
$\eventset\setminus\initeventset=\set{\event_1,\ldots,\event_\nn}$,
i.e.,
each non-initializer event corresponds exactly to one label in $\run$.
%\\
%$\bullet$
\item
If $\lbl_\ii=\tuple{\thread_\ii,\wtype,\xvar_\ii,\val_\ii}$,
then 
$\event_\ii=\tuple{\id_\ii,\thread_\ii,\wtype,\xvar_\ii,\val_\ii}$,
and
if $\lbl_\ii=\tuple{\thread_\ii,\rtype,\xvar_\ii,\val_\ii}$,
then 
$\event_\ii=\tuple{\id_\ii,\thread_\ii,\rtype,\xvar_\ii}$.
An event and the corresponding label
perform identical operations (write or read)
on the same variables.
In the case of a write, they also agree on the written value.
%
%\\
%$\bullet$
\item 
$\id_\ii=\sizeof{
\setcomp{\jj}{(1\leq\jj\leq\ii)\wedge(\thread_\jj=\thread_\ii)}}$.
The identifier of an event shows how it is ordered relative to 
the other events performed by the same thread.
%\\
%$\bullet$
\item
  If $\event_\ii\;[\rf]\;\event_\jj$, then 
$\xvar_\ii=\xvar_\jj$ and $\val_\ii=\val_\jj$.
A read event fetches its value from an event 
that writes the same value to the same variable.
%\\
%$\bullet$
\item
  If $\initx\;[\rf]\;\event_\ii$,
 then  $\val_\ii=0$, i.e., $\event_\ii$ reads the initial value of $\xvar$ 
which is assumed to be $\zeroval$.
 \end{itemize}

\paragraph{Release-Acquire Semantics}
Following~\citet{DBLP:conf/popl/LahavGV16}, we define the Release-Acquire semantics by defining the set of runs whose
associated total traces do not contain certain forbidden cycles.

Given a trace $\trace = \tracetuple$,
we define the derived relation
$\fr$ (from-read) by $\fr:=\cup_{\xvar\in\varset}\frx$, where
$\frx:=(\rfx)^{-1}\compose\cox$.
Intuitively, if $\event\;[\frx]\;\event'$ then the write
$\event'$ overwrites the value read by the read $\event$
(since $\event'$ is coherence-order-after the write event from which
$\event$ gets its value).

\begin{definition}
\label{def:ra}
For a  trace $\trace$, let $\trace\!\models\!\ra$ denote that
the relation $\po\!\cup\!\rf\!\cup\!\cox\!\cup\!\frx$ is acyclic
for each $\xvar \in \varset$.
We define 
$
\ssdenotationof\ra\conf\!:=\!
\setcomp{\trace}{\exists\run\in\runs(\conf).\;
  \run\models\trace\land\istotal\trace\land\trace\models\ra}$, 
i.e.,
$\ssdenotationof\ra\conf$ is the set of total traces generated
under   RA  from a given configuration $\conf$.
%\qed
\end{definition}

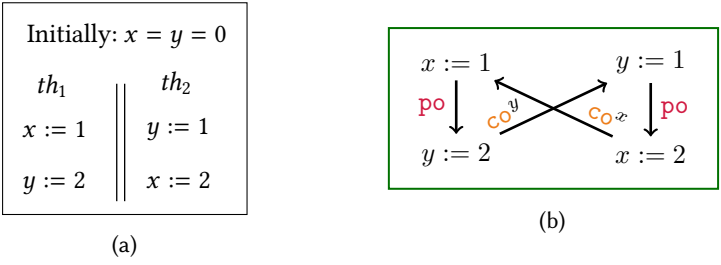
\begin{figure}[tb]
 \begin{minipage}[c]{0.25\linewidth}
  \center
  \subcaptionbox{\label{2+2W:prog:fig}}{%
         
      \begin{tikzpicture}[line width=1pt,framed]
	\node[name=n0] at (0,0) {Initially: $\xvar=\yvar=0$};

	\node[name=n11,anchor=west] at ($(n0.east)+(-85pt,-35pt)$) {$\xvar:=1$};
	\node[name=n12,anchor=north west] at ($(n11.south west)+(0pt,-5pt)$) {$\yvar:=2$};

	\node[name=n21,anchor=west] at ($(n11.east)+(15pt,0pt)$) {$\yvar:=1$};
	\node[name=n22,anchor=north west] at ($(n21.south west)+(0pt,-4pt)$) {$\xvar:=2$};

	\node[name=th1,anchor=south west]at ($(n11.north west)+(5pt,1pt)$) {$\thread_1$};
	\node[name=th2,anchor=south west]at ($(n21.north west)+(5pt,1pt)$) {$\thread_2$};

	\draw[line width= 0.5pt] ($(n11.north east)+(7.5pt,10pt)$) -- ($(n11.south east)+(7.5pt,-20pt)$);
	\draw[line width= 0.5pt] ($(n11.north east)+(10.5pt,10pt)$) -- ($(n11.south east)+(10.5pt,-20pt)$);

	\end{tikzpicture}
  }
  \end{minipage}
  \hspace{10mm}
  \begin{minipage}[c]{0.4\linewidth}
  \center
  \subcaptionbox{\label{2+2W-ss:fig}}{%
      \includegraphics{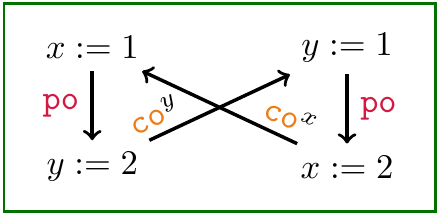}
  }
  \end{minipage}
%  \vspace{-2mm}
  \caption{ (a) The program 2+2W and (b) 
    a Shasha-Snir trace.}
\end{figure}

To illustrate the RA semantics, Figure~\ref{2+2W:prog:fig}
shows a simple program, known as \texttt{2+2W}~\cite{alglaveMT14},
with two threads
$\thread_1$ and $\thread_2$
that communicate
through
two shared variables $\xvar$ and $\yvar$.
Thread
$\thread_1$ writes $1$ to $\xvar$ and then $2$ to $\yvar$.
Symmetrically,
$\thread_2$ writes $1$ to $\yvar$ and then $2$ to $\xvar$.
We would like to check whether 
the writes
$\xvar:=2$ and $\yvar:=2$  can be placed before the writes
$\xvar:=1$ and $\yvar:=1$
in the corresponding coherence order relation ($\co$).
Figure~\ref{2+2W-ss:fig} gives the corresponding trace $\trace_1$.
To improve readability,  we  use a simplified notation for events.
More precisely, we represent an event in a trace by the corresponding program instruction.
For instance, we  write $\xvar:= 1$ instead of $\tuple{1,\thread_1,\wtype,\xvar,1}$.
We observe that  the read-from relation $\rf$ and the from-read relation $\fr$ are empty in $\trace_1$.
Since the relations $\po\cup\rf\cup\co^{\xvar}\cup\fr^{\xvar}$
and $\po\cup\rf\cup\co^{\yvar}\cup\fr^{\yvar}$ are acylic,
it follows by Definition~\ref{def:ra},
that $\trace_1\!\models\!\ra$.

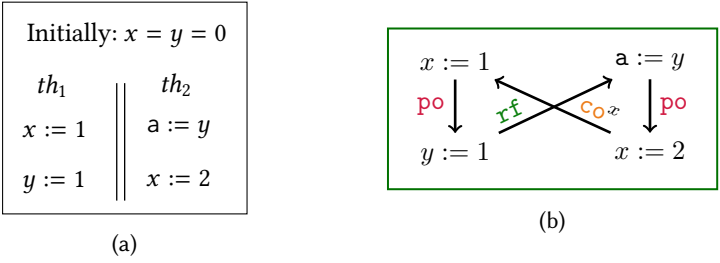
\begin{figure}[tb]
 \begin{minipage}[c]{0.25\linewidth}
  \center
  \subcaptionbox{\label{S:prog:fig}}{%
         
      \begin{tikzpicture}[line width=1pt,framed]
	\node[name=n0] at (0,0) {Initially: $\xvar=\yvar=0$};

	\node[name=n11,anchor=west] at ($(n0.east)+(-85pt,-35pt)$) {$\xvar:=1$};
	\node[name=n12,anchor=north west] at ($(n11.south west)+(0pt,-5pt)$) {$\yvar:=1$};

	\node[name=n21,anchor=west] at ($(n11.east)+(15pt,0pt)$) {$\avar:=\yvar$};
	\node[name=n22,anchor=north west] at ($(n21.south west)+(0pt,-4pt)$) {$\xvar:=2$};

	\node[name=th1,anchor=south west]at ($(n11.north west)+(5pt,1pt)$) {$\thread_1$};
	\node[name=th2,anchor=south west]at ($(n21.north west)+(5pt,1pt)$) {$\thread_2$};

	\draw[line width= 0.5pt] ($(n11.north east)+(7.5pt,10pt)$) -- ($(n11.south east)+(7.5pt,-20pt)$);
	\draw[line width= 0.5pt] ($(n11.north east)+(10.5pt,10pt)$) -- ($(n11.south east)+(10.5pt,-20pt)$);

	\end{tikzpicture}
  }
  \end{minipage}
  \hspace{10mm}
  \begin{minipage}[c]{0.4\linewidth}
  \center
  \subcaptionbox{\label{S-ss:fig}}{%
      \includegraphics{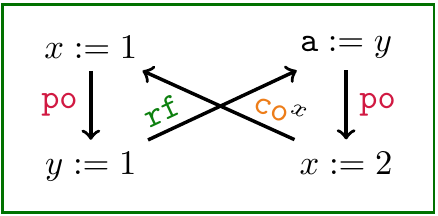}
  }
  \end{minipage}
%  \vspace{-2mm}
  \caption{ (a) The program S and (b) 
    a Shasha-Snir trace.}
\end{figure}

To see the role of read-from relations
in the RA semantics, Figure~\ref{S:prog:fig}
gives another program, known as \texttt{S}~\cite{alglaveMT14}.
In a similar manner to \texttt{2+2W}, the program \texttt{S} has two threads
$\thread_1$ and $\thread_2$
that communicate
through
two shared variables $\xvar$ and $\yvar$.
Thread $\thread_1$
writes $1$ to both $\xvar$ and $\yvar$.
Thread $\thread_2$
reads from $\yvar$ and stores the value to a local register $\avar$
and then writes $2$ to $\xvar$.
We would like to check whether it is possible 
to place the write
$\xvar:=2$  before the write
$\xvar:=1$ in coherence order ($\co$)
and  make
$\avar:=\yvar$ 
read from the write $\yvar:=1$ in the read-from relation ($\rf$).
Figure~\ref{2+2W-ss:fig} gives the corresponding trace $\trace_2$.
Since the relation
$\po\cup\rf\cup\co^{\xvar}\cup\fr^{\xvar}$
is cyclic, it follows by Definition~\ref{def:ra},
that $\trace_2\!\not\models\!\ra$.

Since the result computed by a run is uniquely determined by its
associated trace, we can analyze
a concurrent program under the RA semantics by exploring runs and their
associated traces until all traces in $\ssdenotationof\ra\conf$ have
been generated. One can even define an algorithm which is optimal in that it
explores each trace in $\ssdenotationof\ra\conf$ exactly once~\cite{KLSV:popl18}.
However, it is possible to be more efficient than this, by exploiting the
observation that the results computed in a run, including, e.g., the outcome
of assert statements, depend only on the  program order ($\po$) and
read-from ($\rf$) relations, irrespective of the $\co$ relation. In fact,
the sequence of local states of a thread and the values it writes to shared
variables are affected only by the sequence of values it reads from shared
variables. Hence two runs, whose associated traces have the same $\po$ and
$\rf$ relations, can be regarded as equivalent. An algorithm based on the set
$\ssdenotationof\ra\conf$ will in general be redundant, in that
it explores several traces that are equivalent in this sense.

\paragraph{Weak Traces}
The above observation suggests to base the analysis on a weaker notion of trace.
Define a {\it weak trace} to be a trace whose $\co$ relation is empty.
For a trace $\trace=\tuple{\eventset,\po,\rf,\co}$, we 
define its {\it weakening} by
$\weakof\trace:=\tuple{\eventset,\po,\rf,\emptyset}$, obtained
by removing all coherence edges.
We define the set 
$\wdenotationof\ra\conf:=
\setcomp{\weakof{\trace}}{\trace \in\ssdenotationof\ra\conf}$.
In other words, $\wdenotationof\ra\conf$ is the set of
weakenings of the traces in the total semantics.
Let us introduce a term for such weakenings.
%% traces that can be obtained by weakening some trace in $\ssdenotationof\ra\conf$.
%
For traces
 $\trace=\tracetuple$ and
$\trace'=\tuple{\eventset',\po',\rf',\co'}$, let
$\trace \tordering \trace'$ denote that
$\eventset = \eventset'$, $\po = \po'$, $\rf = \rf'$ and 
$\cox\subseteq(\co')^\xvar$ for
each $\xvar \in \varset$.
We say that a (partial) trace $\trace$ is {\em RA-consistent} if there is a
total trace $\trace'$ with $\trace \refinedby \trace'$ such that
$\trace'\models\ra$.
%
%% Intuitively, a trace is {\it RA-consistent} if 
%% it can be extended to total trace that do not contain any cycles
%% that are forbidden by the RA semantics.
%
In particular, a total trace $\trace$ is RA-consistent if and only
if $\trace\models\ra$. Also,
$\wdenotationof\ra\conf$ is the set of RA-consistent weak traces of $\conf$.

We can now analyze a program by exploring runs until all
weak traces in $\wdenotationof\ra\conf$ have been generated.
In the next sections, we will present such an analysis algorithm, which is
also \emph{optimal} in the sense that it explores each trace in
$\wdenotationof\ra\conf$ exactly once. Such an algorithm
must overcome the challenges of
\begin{inparaenum}[(i)]
\item
  \emph{consistency}, i.e.,
examining only RA-consistent weak traces 
(note that many weak traces are not RA-consistent;
e.g., for the program in Figure~\ref{overview:prog:fig}, letting
both threads read from the write of the other thread is not possible under the RA semantics),
and
\item
\emph{non-redundancy}, i.e.,
to generate each weak trace only once, thereby avoiding unnecessary explorations.
\end{inparaenum}
We solve the consistency challenge in~\cref{saturated:section}, by
defining a {\em saturated semantics}, equivalent with the standard
one, in which runs are associated with {\em partial} traces that
contain precisely those coherence edges that must be present in any
corresponding total trace.
%% , and also makes it easy to check and maintain RA-consistency.
Based on the saturated semantics, 
the non-redundancy challenge is solved by the design of our DPOR algorithm
in~\cref{dpor:section}.

%% A challenge that faces such an algorithm is that a naive approach, which
%% enumerates all combinations of $\po$ and $\rf$ relations, will not work, since
%% many of them are not RA-consistent. For instance, for the program
%% in Fig.~\ref{overview:prog:fig}, the $\rf$ relation where
%% both threads read from the write of the other thread is not possible
%% in an RA-consistent trace.

\section{Saturated Semantics}
\label{saturated:section}
In this section, we address the challenge of exploring only RA-consistent
trace by defining a new semantics for RA, called the {\it saturated semantics}.
The saturated semantics is the basis for our
DPOR algorithm in~\cref{dpor:section},
which generates precisely the weak traces in $\wdenotationof\ra\conf$
for a given configuration $\conf$.

The saturated semantics solves the consistency challenge by
making it easy to  maintain RA-consistency.
We define the semantics in two steps.
First, in~\cref{saturated:trace:section},
we define the notion of a  {\it saturated trace} as
a partial trace which extends a weak trace, whose
(partial) coherence relation contains precisely the edges that
occur in all RA-consistent total extensions of that weak trace.
%
%As a key theorem (Theorem~\ref{saturated:consistent:theorem}),
We show that a saturated trace is RA-consistent
iff it does not contain a cycle that is forbidden by the RA semantics (cf.\ Theorem~\ref{saturated:consistent:theorem}).
Then, 
in~\cref{saturated:semantics:section},
%based on this theorem,
we present two efficient operations that allow adding a new write
(resp.\  read) event to a trace while preserving 
both saturation and RA-consistency.
We use
these operations as the basis to define  our saturated semantics.
Finally,
we show a key theorem (cf.\ Theorem~\ref{ss:partial:semantics:theorem}) that the saturated semantics coincides with the RA semantics
on weak traces.

\subsection{Saturated Traces}
\label{saturated:trace:section}

\begin{definition}
A trace $\trace$ is 
{\it saturated} if, for all variables $\xvar\in\varset$, whenever
$\event,\event' \in \eventset^{\wtype,\xvar}$ with $\event \neq \event'$
such that 
$\event\;[\po\cup\rf\cup\cox]^+\;\event''$ and
$\event'\;[\rfx]\;\event''$, then
$\event\;[\po\cup\rf\cup\cox]^+\;\event'$.
\end{definition}
To motivate this definition, note that
whenever $\event\;[\po\cup\rf\cup\cox]^+\;\event''$ and
$\event'\;[\rfx]\;\event''$ then (under RA) any coherence edge between
$\event$ and $\event'$ must (if present) be directed from
$\event$ to $\event'$; otherwise
$\event'\;[\cox]\;\event$ and $\event'\;[\rfx]\;\event''$ would imply
$\event''\;[\frx]\;\event$, implying
$\event\;[\po\cup\rf\cup\cox\cup\frx]^+\;\event$
(by $\event\;[\po\cup\rf\cup\cox]^+\;\event''\;[\frx]\;\event$),
thereby violating the RA semantics.

The following theorem shows an essential property of
saturated traces, namely that if such a trace does not contain
cycles violating the RA semantics
then it is RA-consistent.

\begin{theorem}
\label{saturated:consistent:theorem}
For a partial trace $\trace$, if $\trace$ is saturated and
$\trace\models\ra$, then $\trace$ is RA-consistent.
\end{theorem}
\begin{proof}
Assume that
$\trace=\tuple{\eventset,\po,\rf,\co}$ is saturated and
$\trace\models\ra$.
We show that there is a total trace
$\trace'=\tuple{\eventset,\po,\rf,\co'}$ 
such that $\trace\tordering\trace'$
and $\trace'\models\ra$.
The lemma then follows immediately.

We define a sequence of traces
$\trace_0\tordering\trace_1\tordering\trace_2,\ldots$ inductively 
such that 
$\trace_0=\trace$,
$\trace_\ii$ is saturated and $\trace_\ii\models\ra$.
For $\ii>0$, if $\trace_\ii=\tuple{\eventset,\po,\rf,\co_\ii}$ is not total
then
$\trace_{\ii+1}$ is derived from 
$\trace_\ii$ by adding the pair $\tuple{\event_\ii,\event'_\ii}$
to the coherence order,
where $\event_\ii,\event'_\ii\in\eventset^{\wtype,\xvar}$
for some $\xvar\in\varset$ with 
$\neg(\event'_\ii\;[\po\cup\rf\cup\cox_\ii]^+\;\event_\ii)$ and
$\neg(\event_\ii\;[\cox_\ii]\;\event'_\ii)$.
Such events $\event_\ii$ and $\event'_\ii$ exist since $\trace_\ii$ is not total
and since $\trace_\ii\models\ra$ by the induction hypothesis.
Also, by the induction hypothesis we know that $\trace_\ii$ is saturated.
It follows that
$\trace_{\ii+1}$ is saturated.
Since $\trace_{\ii}\models\ra$ it follows
that $\trace_{\ii+1}\models\ra$.
Notice that, by construction, we have $\sizeof{\trace_\ii}<\sizeof{\trace_{\ii+1}}$.
It follows that 
there is a $\jj$ such that $\trace_\jj$ is total.
Define $\trace':=\trace_\jj$.
This concludes the proof the the lemma.
\end{proof}

\subsection{Saturated Semantics}
\label{saturated:semantics:section}
Next, we introduce two notions 
that are relevant when adding a new read event to a trace.

\subsubsection{Readability and Visibility}
\label{readbility:visibility:section}
(i)
{\it Readability} identifies
the write events $\event'$  from which read event
 $\event$ can fetch its value, and
{\it visibility} is used to add
new coherence-order edges that are implied by the fact that
the new event $\event$ reads from $\event'$.
Below, we fix a trace $\trace=\tracetuple$.

\begin{figure}[tb]
 \begin{minipage}[c]{0.4\linewidth}
  \center
  \subcaptionbox{\label{add:read:fig}}{%
         
     \begin{tikzpicture}[background rectangle/.style={rounded corners,line width=0.5pt,draw,fill=gray!5}]

\node[inner sep=2pt,circle,draw,fill=myviolet,name=n11] {};
\node[name=l11,anchor=south west,inner sep=0pt] at (n11.north east) {$\event_3$};
\node[inner sep=2pt,circle,draw,fill=myviolet,anchor=east,name=n12] 
at ($(n11.west)+(-35pt,0pt)$) {};
\node[name=l12,anchor=south east,inner sep=0pt] at (n12.north west) {$\event_2$};
\node[inner sep=2pt,circle,draw,fill=myviolet,anchor=north,name=n21] 
at ($(n11.south)+(0pt,-20pt)$) {};
\node[name=l21,anchor=west,inner sep=0pt] at ($(n21.east)+(1pt,0pt)$) {$\event_4$};
\node[inner sep=2pt,circle,draw,fill=myviolet,anchor=north,name=n31] 
at ($(n21.south)+(0pt,-20pt)$) {};
\node[name=l31,anchor=west,inner sep=0pt] at ($(n31.east)+(1pt,0pt)$) {$\event_5$};
\node[inner sep=2pt,circle,draw,fill=myviolet,anchor=north,name=n41] 
at ($(n31.south)+(0pt,-30pt)$) {};
\node[name=l41,anchor=west,inner sep=0pt] at ($(n41.east)+(1pt,0pt)$) {$\event_6$};

\node[inner sep=2pt,circle,draw,fill=myviolet,anchor=east,name=n42]
at ($(n41.west)+(-35pt,0pt)$) {};
\node[name=l42,anchor=east,inner sep=0pt] at ($(n42.west)+(-1pt,0pt)$) {$\event_1$};

\draw[->,line width=1pt] (n12) to node[above]{$\po,\cox$}(n11);
\draw[->,line width=1pt] (n11) to node[right]{$\rf$}(n21);
\draw[->,line width=1pt] (n21) to node[right]{$\po$}(n31);
\draw[->,line width=1pt,dotted] (n31) to node[right]{$\po$}(n41);
\draw[->,line width=1pt,dotted] (n42) to node[below]{$\rfx$}(n41);
\draw[->,line width=1pt,dotted] (n11) to node[above,sloped]{$\cox$}(n42);

\begin{scope}[on background layer]
	\node (nl12) [align=center] at ($(l12.north west)+(-2pt,3pt)$) {};
	\node (nl11) [align=center] at ($(l11.north east)+(2pt,3pt)$) {};
	\node (nl21) [align=center] at ($(l21.south east)+(2pt,0pt)$) {};
	\node (nl31) [align=center] at ($(l31.south east)+(1pt,-2pt)$) {};
	\node (nl42) [align=center] at ($(l42.south east)+(-0pt,-5pt)$) {};
	\node (m1) [align=center] at ($(nl42.north)+(-10pt,40pt)$) {};
	\node (m2) [align=center] at ($(nl42.north)+(40pt,30pt)$) {};
	\node (m3) [align=center] at ($(nl42.north west)+(-15pt,10pt)$) {};

	\draw[bend left, fill=mygray]
    		(nl42.east)
		to (m3)
		to (m1)
      		to (nl12)
      		to (nl11)
      		to (nl21)
      		to (nl31)
		to (m2)
      		to (nl42.east);
      
\end{scope}

\end{tikzpicture} 
 }
  \end{minipage}
  \hspace{5mm}
  \begin{minipage}[c]{0.4\linewidth}
  \center
  \subcaptionbox{\label{add:write:fig}}{%
      \begin{tikzpicture}[background rectangle/.style={rounded corners,line width=0.5pt,draw,fill=gray!5}]

\node[inner sep=2pt,circle,,fill=myviolet,draw,name=n11] {};
\node[name=l11,anchor=south west,inner sep=0pt] at (n11.north east) {$\event_2$};
\node[inner sep=2pt,circle,draw,fill=myviolet,anchor=east,name=n12] 
at ($(n11.west)+(-35pt,0pt)$) {};
\node[name=l12,anchor=south east,inner sep=0pt] at (n12.north west) {$\event_1$};
\node[inner sep=2pt,circle,draw,fill=myviolet,anchor=north,name=n21] 
at ($(n11.south)+(0pt,-25pt)$) {};
\node[name=l21,anchor=west,inner sep=0pt] at ($(n21.east)+(1pt,0pt)$) {$\event_3$};
\node[inner sep=2pt,circle,draw,fill=myviolet,anchor=north,name=n31] 
at ($(n21.south)+(0pt,-25pt)$) {};
\node[name=l31,anchor=west,inner sep=0pt] at ($(n31.east)+(1pt,0pt)$) {$\event_4$};
\node[inner sep=2pt,circle,draw,fill=myviolet,anchor=north,name=n41] 
at ($(n31.south)+(0pt,-30pt)$) {};
\node[name=l41,anchor=west,inner sep=0pt] at ($(n41.east)+(1pt,0pt)$) {$\event_5$};

\begin{scope}[on background layer]
	\node (nl31) [align=center] at ($(l31.south east)+(0pt,-3pt)$) {};
	\node (nl21) [align=center] at ($(l21.south east)+(3pt,-5pt)$) {};
	\node (nl11) [align=center] at ($(l11.north east)+(5pt,3pt)$) {};
	\node (nl12) [align=center] at ($(l12.north west)+(-0pt,8pt)$) {};
	
	%\node (nl42) [align=center] at ($(l42.south west)+(-0pt,-5pt)$) {};
	\node (m1) [align=center] at ($(nl31.north west)+(-55pt,30pt)$) {};
	\node (m2) [align=center] at ($(nl31.west)+(-35pt,-5pt)$) {};

	%\node (m2) [align=center] at ($(nl42.north)+(45pt,30pt)$) {};

	\draw[bend left, fill=mygray]
    		(nl31.south west)
		to (m2)
		to (m1)
      		to (nl12)
      		to (nl11)
      		to (nl21)
      		to (nl31.south west);
      
\end{scope}

\draw[->,line width=1pt] (n12) to node[above]{$\po,\cox$}(n11);
\draw[->,line width=1pt] (n11) to node[right]{$\rf$}(n21);
\draw[->,line width=1pt] (n21) to node[right]{$\po$}(n31);
\draw[->,line width=1pt,dotted] (n31) to node[right]{$\po$}(n41);

\end{tikzpicture}
  }
  \end{minipage}
  \caption{(a) Adding a read event $\event_6$ and (b) 
    a write event $\event_5$.
    The dotted arrows are added due to the new event.}
\end{figure}
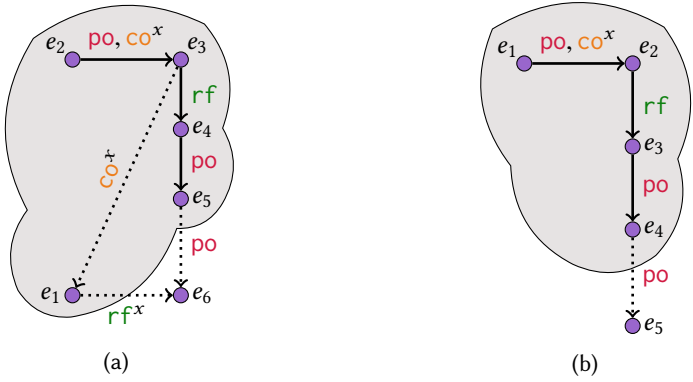

\begin{definition}
For a thread $\thread\in\threadset$ and a variable $\xvar\in\varset$,  
 the set of {\em readable events} $\readable(\trace,\thread,\xvar)$ 
is defined to
be the set of events $\event\in\eventset^{\wtype,\xvar}$ such that
there are no events $\event'\in\eventset^{\wtype,\xvar}$
and $\event''\in\eventset^\thread$
with $\event\;[\po\cup\rf\cup\cox]^+\;\event'$ and
$\event'\;[\po\cup\rf]^*\;\event''$.
\end{definition}
Intuitively, $\readable(\trace,\thread,\xvar)$ contains all write events on
$\xvar$ that are not hidden from thread $\thread$ 
by other write events on $\xvar$.
A new read event on $\xvar$ that is added to $\thread$
can fetch its value from any write event in $\readable(\trace,\thread,\xvar)$.
In fact, if $\thread$ is saturated then 
$\readable(\trace,\thread,\xvar)$ is precisely
the set of write events from which $\event$ can read without
introducing a cycle that is forbidden by the RA semantics.

To illustrate this, let $\trace$ be as in Figure~\ref{add:read:fig}, where the
read  $\event_6$ is about to be added, and let
$\event_2$, $\event_3$, and $\event_1$ be write events on $\xvar$.
The dotted arrows (explained later)
represent edges that will be added to
the traces due to the new event $\event_6$.
Let $\thread$ be the thread of $\event_5$ and $\event_6$.
Then $\event_1,\event_3\in\readable(\trace,\thread,\xvar)$,
while $\event_2\not\in\readable(\trace,\thread,\xvar)$.
Note that reading from $\event_2$ would destroy RA-consistency, since
any corresponding total trace must have $\event_2\;[\cox]\;\event_3$ and hence
$\event_6\;[\frx]\;\event_3$, inducing the cycle
$\event_6\;[\frx]\;\event_3\;[\rf]\;\event_4\;[\po]\;\event_6$
thereby violating the RA semantics.

\begin{definition}
For a thread $\thread\in\threadset$ and a variable $\xvar\in\varset$,  
the {\it visible events} $\visible(\trace,\thread,\xvar)$ is defined to
be the set of events $\event$ in $\readable(\trace,\thread,\xvar)$ such that
there is an $\event'\in\eventset^\thread$ with
$\event\;[\po\cup\rf]^*\;\event'$.
\end{definition}
In other words, the set contains all the readable events on $\xvar$ 
that can reach an event in $\thread$ through $\po$ and $\rf$ edges.
The point of $\visible(\trace,\thread,\xvar)$ is that
if a new read event $\event$ on $\xvar$ is  added to $\thread$, which
reads from an event $\event'$ then
the resulting trace is saturated by adding
a coherence-order edge from each
$\event''\in\visible(\trace,\thread,\xvar)$ to $\event'$.
As illustration, 
in Figure~\ref{add:read:fig} (again before adding $\event_6$), we have
$\event_3\in\visible(\trace,\thread,\xvar)$ 
while
$\event_2\not\in\visible(\trace,\thread,\xvar)$.
This means that if we let $\event_6$ read from 
$\event_1$ then the saturation of the resulting trace adds a coherence-order edge
from $\event_3$ to $\event_1$; on the other hand, no edge is added from
$\event_2$ to $\event_1$.

\begin{figure}
\begin{tikzpicture}[background rectangle/.style={rounded corners,line width=1pt,draw},show background rectangle]
\node[name=premise,rulenode,align=center]{

$\event=\tuple{\sizeof{\eventset^\thread}\!+\!1,\thread,\wtype,\xvar,\val}$,\;\;\;
$\eventset'=\eventset\cup\set\event$,\\
$\po'=\po\cup\setcomp{\tuple{\event',\event}}{\event'\in\eventset^\thread}$,\\
$\trace=\tuple{\eventset,\po,\rf,\co}$,\;\;\;
$\trace'=\tuple{\eventset',\po',\rf,\co}$,
};
\node[name=concl,anchor=north,rulenode] at (premise.south)
{
$\trace\cmovesto{\event}\trace'$
}; 
\draw (premise.south west) -- ($(premise.south east) +(10pt,0pt) $);
\node[anchor=west,text=red,font=\small,color=red,draw,rounded corners,align=left] 
at ($(concl.east)+(2pt,-1pt)$) {{\tt S-TRACE-WRITE}};
\end{tikzpicture}
\begin{tikzpicture}[background rectangle/.style={rounded corners,line width=1pt,draw},show background rectangle]
\node[name=premise,rulenode,align=center]{
$\event=\tuple{\sizeof{\eventset^\thread}\!+\!1,\thread,\rtype,\xvar}$, \;\;\; 
%% $\event=\neweventof{\trace,\thread,\rtype,\xvar}$,
$\eventset'=\eventset\cup\set{\event}$, \;\;\;\;
$\event'\in\readable(\trace,\thread,\xvar)$,\\ 
$\po'=\po\cup\setcomp{\tuple{\event'',\event}}{\event''\in\eventset^\thread}$,\;\;\;
$\rf'=\rf\cup\set{\tuple{\event',\event}}$, \\
$\co'=\co\cup\setcomp{\tuple{\event'',\event'}}
{\event''\in\visible(\trace,\thread,\xvar) \wedge \event'\neq\event''}$\\
 $\trace=\tuple{\eventset,\po,\rf,\co}$,\;\;\;
$\trace'=\tuple{\eventset',\po',\rf',\co'}$,

};
\node[name=concl,anchor=north,rulenode] at (premise.south)
{
$\trace\cmovesto{\tuple{\event,\event'}}\trace'$
}; 
\draw (premise.south west) -- (premise.south east);
\node[anchor=west,text=red,font=\small,color=red,draw,rounded corners,align=left] 
at ($(concl.east)+(16pt,-1pt)$) {{\tt S-TRACE-READ}};
\end{tikzpicture}
%
%\hspace{5mm}
%
%
\begin{tikzpicture}[background rectangle/.style={rounded corners,line width=1pt,draw},show background rectangle]
\node[name=premise,rulenode,align=center]{
$\conf\movesto{\tuple{\thread,\wtype,\xvar,\val}}\conf'$,
$\trace\cmovesto{\event}\trace'$,
$\event=\tuple{\id,\thread,\wtype,\xvar,\val}$
};
\node[name=concl,anchor=north,rulenode] at (premise.south)
{
$\tuple{\conf,\trace}\cmovesto{\event}\tuple{\conf',\trace'}$
}; 
\draw (premise.south west) -- (premise.south east);
\node[anchor=west,text=red,font=\small,color=red,draw,rounded corners] 
at ($(concl.east)+(6pt,0pt)$) {{\tt S-WRITE}};
\end{tikzpicture}
\begin{tikzpicture}[background rectangle/.style={rounded corners,line width=1pt,draw},show background rectangle]
\node[name=premise,rulenode,align=center]{
$\conf\movesto{\tuple{\thread,\rtype,\xvar,\val}}\conf'$,
$\trace\cmovesto{\tuple{\event,\event'}}\trace'$,
$\event=\tuple{\id,\thread,\rtype,\xvar}$,
$\valof{\event'}=\val$
};
\node[name=concl,anchor=north,rulenode] at (premise.south)
{
$\tuple{\conf,\trace}\cmovesto{\tuple{\event,\event'}}\tuple{\conf',\trace'}$
}; 
\draw (premise.south west) -- (premise.south east);
\node[anchor=west,text=red,font=\small,color=red,draw,rounded corners] 
at ($(concl.east)+(26pt,0pt)$) {{\tt S-READ}};
\end{tikzpicture}

%\vspace{-2mm}
\caption{ The saturated transition relation.}
\label{p:traces:fig}
\end{figure}
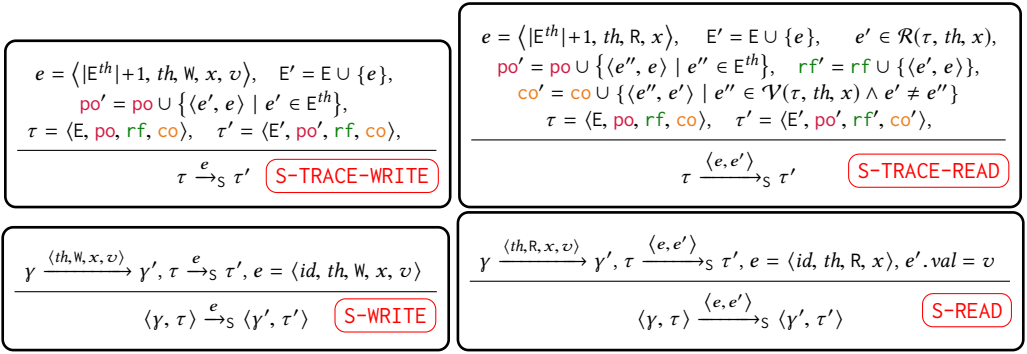

\subsubsection{Saturated Semantics.}
We define the saturated semantics as a transition relation $\cmovesto{}$
on traces (the first two rules of Figure~\ref{p:traces:fig}).
The transition rules correspond to adding
 a new write or read event  to a trace $\trace$
and
obtaining a new trace $\trace'$.
Each transition is labeled by
an {\em observation} $\obs$
which is either a write event $\event$, or
a pair $\tuple{\event,\event'}$, consisting of a 
read event $\event$ that reads from a write event $\event'$.
We define $\eventof\obs:=\event$.
As we will explain below, an important property of the transition relation
is that if the original trace $\trace$ is saturated and has
no cycles forbidden by the RA-semantics, then the
new trace $\trace'$ will satisfy the same conditions.
Therefore, by Theorem~\ref{saturated:consistent:theorem} it follows that
all traces generated according to the saturated
semantics are RA-consistent.
Indeed,
the semantics generates precisely those traces that
are saturated and contain no cycles violating the RA semantics.

Rule {\color{red}{\tt S-TRACE-WRITE}} describes that a saturated trace is extended
with a write event $\event$ by
\begin{inparaenum}[(i)]
\item
adding a new write event $\event$, whose identity
reflects that it is the most recent event performed by its thread,
\item making $\event$ last in the program order of its thread, and
\item
keeping the read-from and coherence relations.
\end{inparaenum}
We observe that
if $\trace$ is saturated and $\trace\models\ra$, then $\trace'$ will
be saturated and $\trace'\models\ra$.
The reason is that $\event$ does not have any successors
w.r.t. $(\po'\cup\rf\cup\co)^+$ in $\trace'$, and hence the only cycles
in $\trace'$ are those that are already in $\trace$.
Since no cycle in $\trace$ violates the RA semantics by assumption,
 no cycle in $\trace'$ will break the RA semantics either.
Figure~\ref{add:write:fig} illustrates 
how to apply
the rule to
extend a  saturated trace with
 a new write event $\event_5$.

Rule {\color{red}{\tt S-TRACE-READ}}  describes that a saturated trace is extended
with a read event $\event$ that reads from a write event $\event'$ by
\begin{inparaenum}[(i)]
\item
adding the new read event $\event$,
\item
ensuring that the write event $\event'$
is readable for the thread $\thread$ w.r.t.\ the variable $\xvar$ in
$\trace$,
\item
extending the program order 
in the same manner as in {\color{red}{\tt S-TRACE-WRITE}},
\item
extending the read-from relation to reflect that $\event$ reads from $\event'$,
and
\item
extending the coherence order by an edge from each visible event to $\event'$.
\end{inparaenum}
If $\trace$ is saturated then $\trace'$ will also be saturated since
we add all the additionally needed coherence-order edges,
namely the ones from the events in $\visible(\trace,\thread,\xvar)$
to $\event'$.
Furthermore, if $\trace$ is saturated and
$\trace\models\ra$ then $\trace'\models\ra$.
%
%% This follows from the definitions of the sets
%% $\readable(\trace,\thread,\xvar)$ and
%% $\visible(\trace,\thread,\xvar)$ 
%% which imply that the coherence-order edges from
%% $\visible(\trace,\thread,\xvar)$
%% to $\event'$ cannot be part of any cycle in $\trace'$.
%
The reason is that  any new cycle violating the
RA semantics in $\trace'$
would include an edge from an event
$\event''\in\visible(\trace,\thread,\xvar)$ to $\event'$.
However, such a cycle implies that
$\event'\;[\po\cup\rf\cup\cox]^+\;\event''$,
i.e.,
$\event'\not\in\readable(\trace,\thread,\xvar)$
which is a contradiction.
Figure~\ref{add:read:fig} illustrates 
how to apply
the rule to
extend a saturated trace %of $\set{\event_1,\cdots,\event_5}$ 
with
 a new read event $\event_6$.

The rules {\color{red}{\tt S-WRITE}} and {\color{red}{\tt S-READ}}
describe how the transition relation $\cmovesto{}$ on saturated traces induces
a corresponding transition relation on pairs of configurations and
saturated traces, in the natural way.
We use $\tuple{\conf,\trace}\cmovesto{}\tuple{\conf',\trace'}$
to denote that
$\tuple{\conf,\trace}\cmovesto\obs\tuple{\conf',\trace'}$
for some $\obs$, and use $\cmovesto*$ to denote the reflexive transitive
closure of $\cmovesto{}$.
We  define 
$\csuccof{\conf,\trace}:=
\setcomp{\tuple{\conf',\trace'}}
{\tuple{\conf,\trace}\cmovesto{}\tuple{\conf',\trace'}}$,
i.e., it is the set of successors of the pair
$\tuple{\conf,\trace}$ w.r.t.\  $\cmovesto{}$.

\subsubsection{Properties of the Saturated Transition Relation}
We describe three properties of the saturated semantics, namely
efficiency, deadlock-freedom, and correctness.

\paragraph{Efficiency.}
The sets $\readable(\trace,\thread,\xvar)$
and $\visible(\trace,\thread,\xvar)$ 
can both be computed in polynomial time.
To see this, suppose we are given a 
$\trace=\tracetuple$, a thread $\thread$, and
a variable $\xvar$.
A polynomial time algorithm for computing the set
$\readable(\trace,\thread,\xvar)$ can be defined consisting of the 
following three steps:
\begin{enumerate}
\item
Compute the transitive closure of the relation $\po\cup\rf\cup\cox$
using, e.g., the Floyd-Warshall algorithm~\cite{Cormen09a}.
%\patodo{Citation needed?}
%
This will cost $O({\sizeof\eventset}^3)$ time.
\item
Compute the set of events 
$\event'$ such that
$\event'\;[\po\cup\rf]^*\;\event''$ for some 
$\event''\in\eventset^\thread$.
This will cost $O({\sizeof\eventset}^2)$ time.
\item
For each event 
$\event\in\eventset^{\wtype,\xvar}$, check whether 
there is an event $\event'$ in the set computed in step (ii)
with $\event\;[\po\cup\rf\cup\cox]^+\;\event'$.
If not, add the event $\event$ to the set 
$\readable(\trace,\thread,\xvar)$.
This will cost $O({\sizeof\eventset}^2)$ time.
\end{enumerate}
The set $\visible(\trace,\thread,\xvar)$
can be computed similarly.
This gives the following lemma.
\begin{lemma}
\label{visible:readable:poly:lemma}
For a trace $\trace$, a thread $\thread$, and a variable $\xvar$,
we can compute the sets 
$\readable(\trace,\thread,\xvar)$ and
$\visible(\trace,\thread,\xvar)$
in polynomial time.
\end{lemma}
By Lemma~\ref{visible:readable:poly:lemma} it follows that
we can compute each step of $\cmovesto{}$ in polynomial
time.
This property is not satisfied by all memory models.
For instance, calculating the successors of a trace in the
SC semantics amounts to solving an NP-complete problem
\cite{DC-DPOR@POPL-18}.

\paragraph{Deadlock-Freedom.}
The saturated semantics is deadlock free in the sense that
if a thread can perform a transition from a configuration then 
there is always a corresponding move in the saturated semantics.
This is captured by the following lemma (which follows immediately from the 
definitions).
\begin{lemma}
\label{saturated:deadlock:lemma}
For a configuration $\conf$ and trace $\trace$, if
$\trace\models\ra$ and 
$\succof\conf\!\neq\!\emptyset$ then 
$\csuccof{\conf,\trace}\neq\emptyset$.
\end{lemma}
%% Lemma~\ref{saturated:deadlock:lemma} by the
%% definition of the saturated semantics (Fig.~\ref{p:traces:fig})
%% as follows.
%% %
%% Let $\trace=\tracetuple$.
%% %
%% If 
%% $\conf\movesto{\thread,\wrtype,\xvar,\val}\conf'$ 
%% we can extend $\trace$ by an event $\event$ as defined in 
%% the rule {\color{red}{\tt S-TRACE-WRITE}}
%% to obtain $\trace'$ and by the rule
%% {\color{red}{\tt S-WRITE}} it follows that
%% $\tuple{\conf,\trace}\cmovesto{\event}\tuple{\conf',\trace'}$.
%% %
%% On the other hand, suppose that
%% $\conf\movesto{\tuple{\thread,\rtype,\xvar,\bullet}}\bullet$, i.e.,
%% the thread $\thread$ performs a read statement on the variable
%% $\xvar$.
%% %
%% Then by the definition of $\visisble$ we know that
%% $\visible(\trace,\thread,\xvar)\neq\emptyset$.
%% \patodo{Do we need $\trace$ to be saturated and RA-consistent?}
%% and hence there some $\event'\in\visible(\trace,\thread,\xvar)$.
%% %
%% We can  extend $\trace$ by a pair $\tuple{\event,\event'}$ 
%% as defined in  the rule {\color{red}{\tt S-TRACE-READ}
%% to obtain $\trace'$ and by the rule
%% {\color{red}{\tt S-READ}} it follows that
%% $\tuple{\conf,\trace}\cmovesto{\event}\tuple{\conf',\trace'}$
%% %

\paragraph{Correctness.}
The following lemma states the fact that the saturated semantics preserves saturation and RA-consistency.

\begin{lemma}
\label{saturated:lemma}
If $\trace$ is saturated,
$\trace\models\ra$, and 
$\trace\cmovesto{\obs}\trace'$
then
$\trace'$ is saturated and $\trace'\models\ra$.
%\qed
\end{lemma}

Define
$\cdenotationof\ra\conf:=
\setcomp{\trace}{\exists\conf'.\;\tuple{\conf,\emptytrace}\cmovesto*\tuple{\conf',\trace}}$, i.e.,
it is the set of traces that can be generated by sequences of
$\cmovesto{}$-transitions, starting from $\conf$ and the empty trace.
The following theorem states that the RA semantics and the saturated
semantics coincide on weak traces, i.e.,
$\cdenotationof\ra\cdot$  and $\ssdenotationof\ra\cdot$
generate the same set of weak traces.

%\phongtodo{the two denotations for total and saturated traces are not really ``equivalent''. We need to fix here}
%\vspace{-2mm}

\begin{theorem}
\label{ss:partial:semantics:theorem}
For any configuration $\conf$, we have
%\vspace{-2mm}
%% $\weakof{\pdenotationof\ra\conf}=\weakof{\ssdenotationof\ra\conf}$
%\[
$\setcomp{\weakof{\trace}}{\trace \in\cdenotationof\ra\conf}
=
\wdenotationof\ra\conf$.
%\;\;\;\;\;
%\qed
%\]
\end{theorem}

\subsection{Example}
\label{saturated:example:section}

We will give an example (Figure~\ref{saturated:semantics:fig})
to illustrate why saturation is important
in the semantics, and how the semantics preserves saturation
of traces.
To that end, we will consider the simple program 
in Figure~\ref{overview:prog:fig} again.
To make the presentation easier to read, 
we will use a simplified notation for events and observations.
First, we  represent an event in a trace
by the corresponding program instruction.
For instance,
 we will write $\xvar:=1$ instead of
$\tuple{1,\thread_1,\wtype,\xvar,1}$, and
we will write $\areg:=\xvar$ instead of
$\tuple{2,\thread_1,\rtype,\xvar}$.
We write an observation $\tuple{\event,\event'}$
as $\avar=2$ if $\event$ corresponds to a read statement of the form
$\areg:=\xvar$ and
$\event'$ corresponds to a write statement of the form
$\xvar:=2$.
In this particular example, we can use this simplified notation
since all the statements
in the program are different.
%% and therefore 
%% each events and observations are
%%  uniquely determined by the corresponding program statements
%% (without the need to mention the event of thread identifiers).
%

\begin{figure}[t]
%\begin{wrapfigure}{r}{.60\textwidth}
%\vspace{-1mm}
\includegraphics{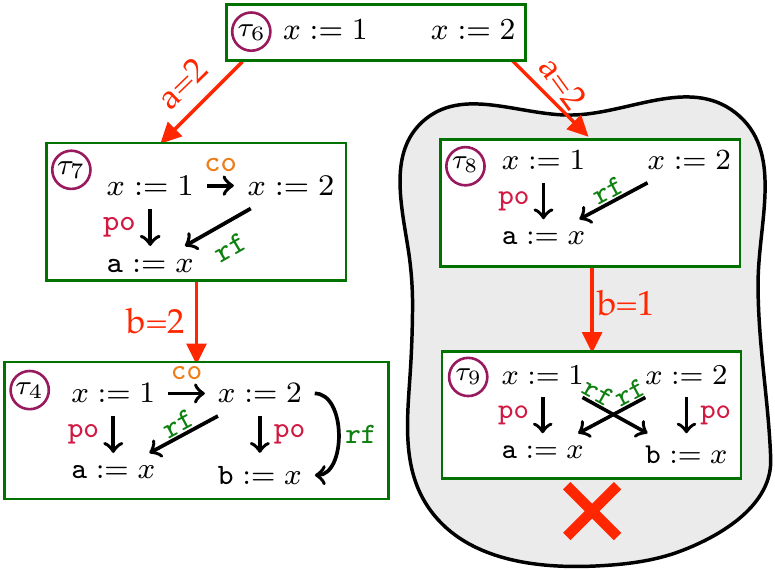}
 \caption{Example illustrating the saturated semantics.}
 \vspace{1mm}
\label{saturated:semantics:fig}
\end{figure}
%\end{wrapfigure}

%\begin{figure}[tb]
Suppose that we have generated the trace $\trace_6$ containing 
the write events $\xvar:=1$  and $\xvar:=2$,
corresponding to $\thread_1$ and $\thread_2$ running
their first instructions respectively.
The trace $\trace_6$ is trivially saturated since it does not contain
any read events.
The event $\xvar:=1$ is visible to the thread $\thread_1$, and
the event $\xvar:=2$ is visible to the thread $\thread_2$.
Both events are readable by both $\thread_1$ and $\thread_2$.
Suppose that $\thread_1$ executes the read instruction
$\areg:=\xvar$
and it chooses to read from the event $\xvar:=2$.
This corresponds to performing the observation
$\areg=2$.
Since $\xvar:=1$ is visible to $\thread_1$ our saturated semantics will
add a $\co$-edge from $\xvar:=1$ to $\xvar:=2$
which means that the resulting trace $\trace_7$ is saturated.
In $\trace_7$, the only readable event by $\thread_2$
is $\xvar:=2$ (since $\xvar:=1$ is now hidden by $\xvar:=2$).
If $\thread_2$ performs the read instruction $\breg:=\xvar$
then it can only read from the event $\xvar:=2$,
corresponding to
 performing the observation
$\breg=2$ and leading to the trace $\trace_4$ 
(the same trace as the one in Figure~\ref{overview:SS:prog:fig}).
Since the only visible event in  $\trace_4$  is 
 $\xvar:=2$, the semantics will not add any new $\co$-edges.
Notice that $\trace_4$ is already saturated.

Next, we will show why saturation is crucial for the semantics.
Let us return to the scenario where we are in $\trace_6$ and
$\thread_1$ performs its read instruction
$\areg:=\xvar$
and it chooses to read from the event $\xvar:=2$.
Suppose that we do not add the $\co$-edge from $\xvar:=1$ to $\xvar:=2$
and thus obtain $\trace_8$ which is not saturated.
If $\thread_2$ performs the read instruction $\breg:=\xvar$ then it cannot
deduce from $\trace_8$ that it cannot read from  $\xvar:=1$.
However, if $\thread_2$ reads from $\xvar:=1$,  we obtain $\trace_9$
which is not RA-consistent.
The reason is that regardless of whether we put a $\co$-edge
from $\xvar:=1$ to $\xvar:=2$ or the opposite, we will obtain 
a  cycle that is forbidden by the RA semantics.

\paragraph{Redundancy.}
While the saturated semantics generates only RA-consistent traces,
it still suffers from the redundancy problem.
More precisely, several runs of the program may produce the same weak trace.
Consider again the simple program
in Figure~\ref{overview:prog:fig}.
Consider two runs, namely $\run_1$
where first $\thread_1$ executes $\xvar:=1$ and 
then $\thread_2$ executes $\xvar:=2$, and
$\run_2$
where first $\thread_2$ executes $\xvar:=2$ and 
then $\thread_1$ executes $\xvar:=1$.
The runs $\run_1$ and $\run_2$ have the same trace, namely $\trace_6$ in
Figure~\ref{saturated:semantics:fig}, and hence
exploring both of them in the analysis would be wasteful.
While this particular scenario is quite simple, 
redundant explorations may have quite
complex forms.
The DPOR algorithm in~\cref{dpor:section} aims
to obtain
an optimal search, i.e., consider only one run per weak trace of the input program.

\section{DPOR Algorithm}
\label{dpor:section}
%\bjcomcom{}

In this section, we present our DPOR algorithm.
For a terminating program, it systematically explores the whole set of
weak traces  that can be generated according to the saturated semantics.
The algorithm is sound, complete, and optimal in the sense that
it produces each weak trace corresponding to a terminated run precisely once.
We achieve optimality by combining the
  saturated semantics with an exploration algorithm
ensuring that  no two generated traces will  have same
program order and read-from relations.
Moreover, the algorithm is deadlock-free in the sense that
the exploration only ends at points when the considered 
program configuration is terminal.
We will first give a detailed description of the algorithm 
in~\cref{description:algorithm:section}, then
provide a complete example that illustrates the main ingredients of the algorithm
in~\cref{dpor:example:section}, 
and finally
state its properties in~\cref{properties:algorithm:section}.

\subsection{Algorithm}
\label{description:algorithm:section}

The DPOR algorithm explores the program
according to the saturated semantics using
the main procedure $\explore$,
starting with an input configuration, a trace, and 
a sequence of observations.
%
%The latter represents the sequence of events that has been considered up to the current point in the exploration.
%

%
For a write event, the algorithm merely adds the event
and continues with the next event.
For a read event $\event$, the algorithm also adds $\event$,
but it continues in several separate branches 
each corresponding to a different write event
from which $\event$ can fetch its value.
Besides, the algorithm must handle the case that $\event$ 
may read from a write event $\event'$ that will only be added to the 
trace later in the exploration.
We say that $\event'$ has been {\it postponed} 
w.r.t.\ $\event$. 
When $\event'$ is eventually generated, 
the DPOR algorithm will detect it and 
{\it swap}  it with $\event$, thus
making  it possible for $\event$ to read from $\event'$.
To swap $\event$ and $\event'$, we also need 
to include the sequence of
observations that are
``necessary'' for generating $\event'$.
We will refer to such a sequence as a {\it schedule} 
(cf.\ the  $\declarepostponed$ procedure,~\cref{declare:postponed:section}).
All the generated schedules will eventually
  be executed thus swapping $\event$ with all the write events
that are postponed w.r.t.\ it
 (cf.\ the  $\rschedule$ procedure,~\cref{rschedule:section}).

\begin{algorithm}[t]
%\small{
\KwIn{
	$\conf$ is a configuration, 
	$\trace=\tracetuple$ is a trace,
	$\obsseq$ is an observation sequence.
	}

\uIf(\tcp*[f]{handle a write event}){$\exists \conf\movesto{\tuple{\thread,\wtype,\xvar,\val}}\conf'$}{\label{explore:if:write:line}
  	{\bf let} $\event$ be $\tuple{\sizeof{\eventset^\thread}\!+\!1,\thread,\wtype,\xvar,\val}$ 
	and $\trace'$ be such that $\tuple{\conf,\trace}\cmovesto\event\tuple{\conf',\trace'}$
	\tcp*{follow  \texttt{S-WRITE}}\label{explore:create:event:line}
  	$\explore(\conf',\trace',\pth\app\event)$\;\label{explore:reccall:line}
  	$\declarepostponed(\trace',\pth\app\event)$\;\label{explore:conflict:line}}

\uElseIf(\tcp*[f]{handle a read event}){$\exists \conf\movesto{\tuple{\thread,\rtype,\xvar,*}}*$}{\label{explore:if:read:line}
 	{\bf let} $\event$ be $\tuple{\sizeof{\eventset^\thread}\!+\!1,\thread,\rtype,\xvar}$\;
  	$\scheduledof{\event}\assigned\emptyset$; $\swapof{\event}\assigned\true$\;\label{explore:initread:line}
  	
	\lFor(\tcp*[f]{follow  \texttt{S-READ}}){$\event',\conf',\trace'$: $\tuple{\conf,\trace}\cmovesto{\tuple{\event,\event'}}\tuple{\conf',\trace'}$}{\label{explore:aval:line}
   		 $\explore(\conf',\trace',\pth\app\tuple{\event,\event'})$\label{explore:if:read:for1:body:line}}
  	
	\lFor{$\schedule\in\scheduledof{\event}$}{\label{explore:if:read:for2:line}
      		$\rschedule(\conf,\trace,\obsseq,\schedule)$}
}
%}
\caption{$\explore(\conf,\trace,\obsseq)$}
\label{dpor:fig}
\end{algorithm}

\subsubsection{$\explore$ - The main procedure   (Algorithm~\ref{dpor:fig})}
\label{explore:section}
The depth-first exploration 
is given in Algorithm~\ref{dpor:fig}.
The $\explore(\conf,\trace,\obsseq)$ procedure 
explores
all RA-consistent weak traces
 of the program
 $\prog$
 that are generated 
 from
 %the recursive procedure $\explore(\conf,\trace,\obsseq)$, which is called when the exploration has reached 
 a
configuration $\conf$ and a saturated trace $\trace$, where
$\conf$ and $\trace$ have been generated by performing
a sequence of observations $\obsseq$.
As we   describe  below, the sequence $\obsseq$ is used for swapping
read events with write events that are postponed w.r.t.\ them.
Initially, the procedure is called with 
an  initial configuration $\conf$, the empty
trace $\emptytrace$, and the empty observation sequence $\emptyword$.

In a call, if $\conf$ has no successors, 
then $\explore$ returns to its caller.
Otherwise,
it considers an enabled write or read statement.
If a write statement is enabled, then one such write statement
is selected non-deterministically, and the corresponding event 
$\event$ is created (at line~\ref{explore:create:event:line}).
This event $\event$  is added to the trace according to the saturated semantics
and also added to the sequence of observations, whereafter
$\explore$ is called recursively to continue the exploration
(at line~\ref{explore:reccall:line}).
After the recursive call has returned, the algorithm
calls   $\declarepostponed$, which  
finds the read events $\event'$ in the input  exploration sequence
$\obsseq$, which would be able to read from the write $\event$ if $\event$
were performed before $\event'$.
For each such a read event $\event'$,
 $\declarepostponed$   creates
a {\it schedule} for $\event'$, which is a sequence of
observations that can be explored from the point where $\event'$ was
performed, to let the write $\event$ occur before $\event'$ so that
$\event'$ can read from $\event$
(see a detailed description in~\cref{declare:postponed:section}).
%\patodo{Last sentence difficult to read}
%

If a read statement is enabled, then a read event $\event$ is created,
and the set $\scheduledof\event$  is initialized to
the empty set (at line~\ref{explore:initread:line}).
The set $\scheduledof\event$ will be gradually updated by the $\declarepostponed$
procedure when subsequent writes are explored.
We also associate a Boolean flag $\swapof\event$ for 
each read event $\event$.
The flag indicates whether $\event$ is {\it swappable},
i.e., whether following write events should consider
$\event$ for swapping or not.
The reason for including this flag is to prevent $\declarepostponed$
from swapping read events that occur in a schedule; this
would lead to redundant explorations and a violation of the optimality
of the algorithm. 
After that, for each already generated write event $\event'$
from which $\event$ can read its value,
$\explore$ is called recursively to continue the exploration
(at line~\ref{explore:if:read:for1:body:line}).
After these  calls have returned, the set of schedules
%that have been 
collected in $\scheduledof\event$
for the read event $\event$
 is considered.
Each such a schedule  is explored by the $\rschedule$ procedure, thereby allowing
$\event$ to fetch its value from  the corresponding write event.

\begin{algorithm}[t]

%\small{
\KwIn{
	$\trace=\tracetuple$ is  a trace and
	$\obsseq$ is an explored observation sequence.
	}

{\bf let} $\event$ be $\lastof\obsseq$ and $\xvar$ be $\varof\event$\;

\For(\tcp*[f]{look for the closest event $\event'$ that has postponed $\event$}){$\ii\assigned \sizeof\obsseq-1$ \KwTo $1$}{\label{declare:postponed:main:loop:line}
  	%{\bf let} $\event'$ be $\event''$ if $\obsseq[\ii]=\event''$ and be $\event''$ if $\obsseq[\ii]=\tuple{\event'',\event'''}$\;
	{\bf let} $\event'$ be $\eventof{\obsseq[\ii]}$\;
  	\If{$\event'\in\eventset^{\rtype,\xvar} \land
		\neg(\event'\;[\po\cup\rf]^+\event) \land
		\swapof{\event'}$}{\label{declare:postponed:find:read:line}
    		$\schedule\assigned\emptyword$\;
   		 \For(\tcp*[f]{get all events after $\event'$ in $\obsseq$ and precedes $\event$ in (\texttt{po}$\cup\rf)^+$}){$\jj\assigned\ii+1$ \KwTo $\sizeof\obsseq-1$}{\label{declare:postponed:dependency:line}
      			{\bf let} $\event''$ be $\eventof{\obsseq[\jj]}$\; 
     			\lIf{$\event''\;[\po\cup\rf]^+\event$}{
          			$\schedule\assigned\schedule\app\obsseq[\jj]$
       			}
		}
    		\If{$\nexists\schedule'\in\scheduledof{\event'}.\;\schedule'\obsseqequiv{\schedule\app\event\app\tuple{\event',\event}}$}{\label{checking:line}
          		$\scheduledof{\event'}\assigned
				\scheduledof{\event'}\cup\set{\schedule\app\event\app\tuple{\event',\event}}$
			\tcp*{allow $\event'$ to read from $\event$}			
			\label{declare:postponed:revent:line}
       		}
		%$\mbox{\qquad \qquad \quad if no equivalent schedule is in }\scheduledof{\event'}$\;
          \textbf{break}\;
  }
}
%}
\caption{$\declarepostponed(\trace,\obsseq)$}
\label{declare:postpone:fig}
\end{algorithm}
%\vspace{-5mm}
%\phongtodo{have changed lines 2 and 11 in Alg 2}

\subsubsection{The $\declarepostponed$ procedure (Algorithm~\ref{declare:postpone:fig})}
\label{declare:postponed:section}
This procedure inputs a trace $\trace$ (the trace that has been 
built up to this point) and a sequence of  observations 
$\obsseq$ whose last element is a write event $\event$.
The algorithm  
finds the closest $\event'$ to $\event$ in $\obsseq$
for which $\event$ can be considered as a postponed write.
The algorithm  then adds
a schedule to the set $\scheduledof{\event'}$ %in order to 
to
allow 
$\event'$ to read from $\event$ in a new exploration.
%
% In other words, a schedule is a sequence of events to be run one after the other,  that will trigger the reading of $\event$ by $\event'$.
%
The criterion for considering such a read event $\event'$
is that it satisfies four conditions, namely
\begin{inparaenum}[(i)]
\item
  $\event'$ reads from the same variable to which $\event$ writes,
\item
$\event'$  does not precede 
  $\event$ w.r.t.\ the relation $(\po\cup\rf)^+$, 
  \item
    $\event'$ is swappable, and
    \item $\event'$ is closest to $\event$.
\end{inparaenum}
The first three conditions are checked at line~\ref{declare:postponed:find:read:line}.
The  condition (iv) is satisfied by the break statement at line 11
which makes  the procedure stop after finding the first read event 
satisfying the first three conditions.

After finding such a read event $\event'$, 
a schedule $\schedule$ is created. 
The schedule consists of all the events
following $\event'$ in $\obsseq$ and which precede
 $\event$ w.r.t.\ the relation $(\po\cup\rf)^+$ (at line~\ref{declare:postponed:dependency:line}). 
The schedule $\schedule$
has
the write event $\event$ and then the pair $\tuple{\event',\event}$ at the end,
thereby achieving the goal of making $\event'$ read from $\event$
(at line~\ref{declare:postponed:revent:line}).

The new schedule $\schedule$ is added to the set $\scheduledof{\event'}$ only 
if the  latter does not already contain a schedule  $\schedule'$ which
is {\it equivalent} to $\schedule$.
We consider $\schedule$ and
$\schedule'$ to be equal, denoted $\schedule\obsseqequiv\schedule'$,
if they contain the same set
of observations (possibly in different orders).
Notice that if $\schedule\obsseqequiv\schedule'$ and we execute $\schedule$ or $\schedule'$ then we reach identical
traces which means that $\schedule$ need not be added
to the set $\scheduledof{\event'}$ if $\schedule'$ is already in $\scheduledof{\event'}$,
and that having  both  $\schedule$ and  $\schedule'$ in $\scheduledof{\event'}$
would in fact violate the optimality condition.

\begin{algorithm}[t]
\SetAlgorithmName{Algorithm}{}{}
%\small{
\KwIn{	$\conf$ is a configuration, 
		$\trace$ is a trace, 
		$\obsseq$ is  an explored-observation sequence, 
		and $\schedule$ is a schedule.}

\uIf(\tcp*[f]{explore the sequence of observations one by one}){$\schedule\neq\emptyword$}{

	{\bf let} present $\schedule$ in the form $\obs\app\schedule'$ and
	%{\bf let} $\event$ be $\event^{\wtype}$ if $\obs=\tuple{\event^{\wtype},\event^{\rtype}}$ and be $\obs$ otherwise\;
	%{\bf let} 
	$\event$ be $\eventof{\obs}$\;
	{\bf choose} $\conf',\trace': \tuple{\conf,\trace}\cmovesto\obs\tuple{\conf',\trace'}$
	\tcp*{follow \texttt{S-WRITE} and \texttt{S-READ}} \label{rschedule:execute:line}
	\lIf{$\typeof\event=\rtype$}{
		$\swapof{\event}\assigned\false$\label{rschedule:non:swap:line}
	}
	$\rschedule(\conf',\trace',\obsseq\app\obs,\schedule')$\label{rschedule:recursion:line}\;
}
\lElse{$\explore(\conf,\trace,\obsseq)$\label{rschedule:explore:line}}
%}
\caption{$\rschedule(\conf,\trace,\obsseq,\schedule)$}
\label{rschedule:fig}
\end{algorithm}
%\phongtodo{have changed line 4 in Alg 3}

\subsubsection{The $\rschedule$ procedure (Algorithm~\ref{rschedule:fig})}
\label{rschedule:section}
The procedure inputs a configuration $\conf$, a trace $\trace$,
a sequence of observations $\obsseq$, and a schedule $\schedule$.
It explores the sequence of scheduled observations in $\schedule$ one by one, 
by calling itself recursively (at line~\ref{rschedule:recursion:line}), after which $\explore$ is called to continue
the exploration beyond the end of $\obsseq$ (at line~\ref{rschedule:explore:line}).
Note that each
observation corresponds to a statement which by construction is enabled.
During this exploration,
read events
%, except the last one, 
%\phongtodo{have updated}
are declared non-swappable by setting the corresponding
entry in $\swap$ to false.

\subsection{Complexity}
Our DPOR algorithm spends \emph{polynomial time} for each explored trace. 
This complexity is estimated using 
the following facts.
\begin{enumerate}
 \item Each single schedule is at most \emph{linear} (in the size of the program), since its length is bounded by the length of the program. 
 \item
  It is indeed possible (in pathological cases) that the exploration will produce an exponential number of schedules. However, each of these schedules will result in a different weak trace; therefore this happens only when the number of weak traces of the program is exponential. Thus (importantly) the effort in constructing these schedules is \emph{never wasted}. 
 \item
  Even if the number of schedules is exponential, it can be checked in \emph{polynomial} (at most quadratic) time whether a new schedule is equivalent to an existing one. This can be done by organizing the schedules into a tree and systematically compare the new schedule from the beginning to the schedules in the tree (the essential operation is to compare read-from edges). 
\end{enumerate}

\subsection{Example}
\label{dpor:example:section}

\begin{figure}[tb]

\includegraphics[width=0.8\textwidth]{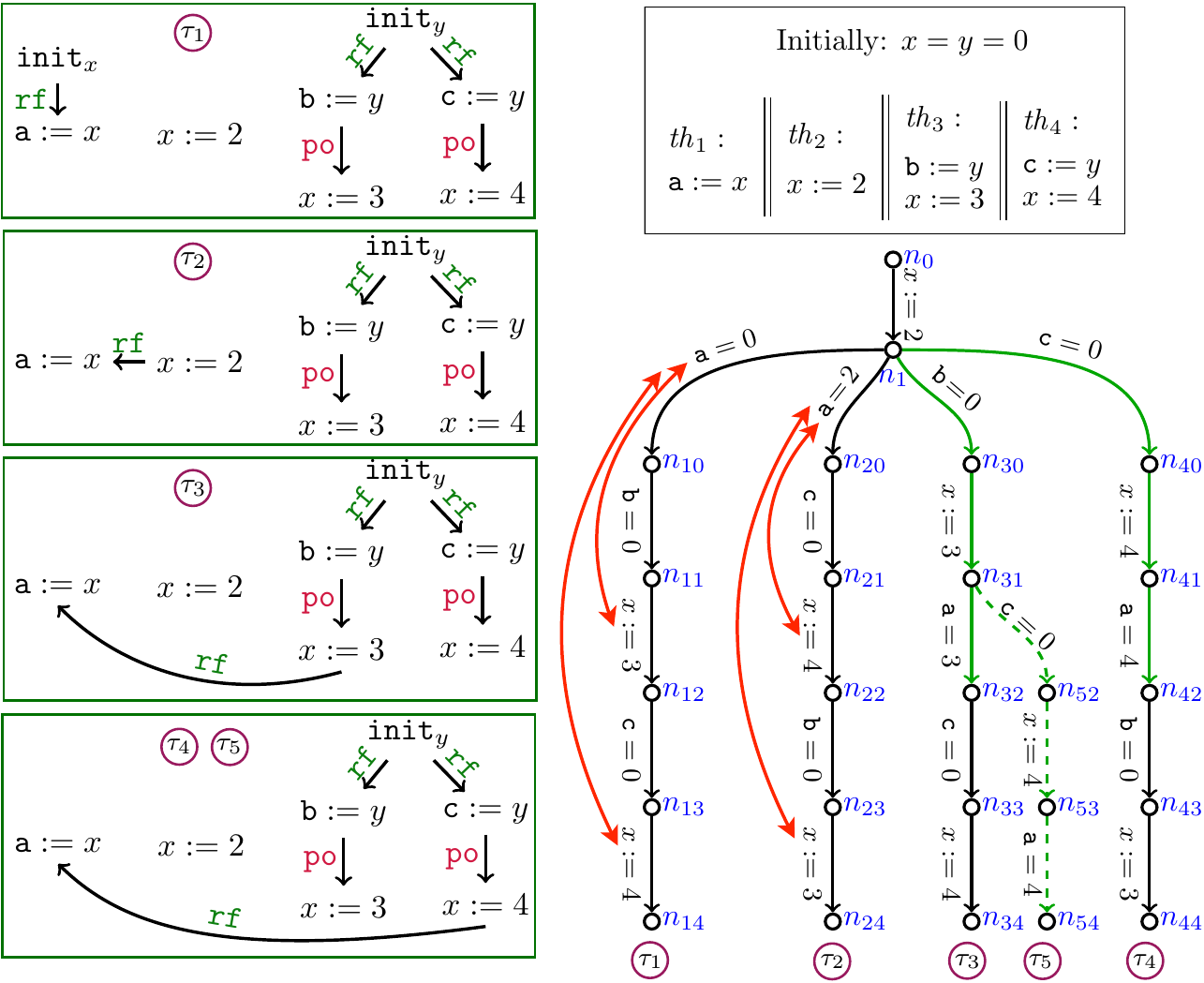}
  \caption{ Example illustrating the DPOR algorithm.}
\label{dpor:example:fig}
\end{figure}

Figure~\ref{dpor:example:fig} illustrates the recursion tree corresponding to
a run of the DPOR algorithm
on a simple program consisting of four threads 
$\set{\thread_1,\thread_2,\thread_3,\thread_4}$ and two shared variables $\set{\xvar,\yvar}$.
We will use the same simplified notation 
for events and observations
as in the example in~\cref{saturated:example:section}.

The  nodes in the tree represent the recursive calls of $\explore$.
The node $\tnode_0$ represents the first call with an empty trace
and an empty input observation sequence.
% \patodo{how about initial configuration?}
%
From $\tnode_0$,  $\explore$  selects the only available write statement
$\xvar:=2$ (of $\thread_2$), making a recursive call and moving to $\tnode_{1}$
(line~\ref{explore:if:write:line} of $\explore$).
At node $\tnode_1$, one of the three read statements,
namely $\areg:=\xvar$ is selected
(line~\ref{explore:if:read:line} of $\explore$).
There are two possible write events from which 
the event $\areg:=\xvar$  can read, namely the initial write event 
$\initx$ which gives the value $0$ and the event $\xvar:=2$
which gives the value $2$.
Accordingly, when $\explore$ runs the for-loop at line~\ref{explore:aval:line},
it will eventually create two branches leading to 
the nodes $\tnode_{10}$ and $\tnode_{20}$.
In the first iteration of the for-loop, a recursive call is made leading to the node $\tnode_{10}$ from where 
the statement $\breg:=\yvar$ is selected.
There is one write event from which $\breg:=\yvar$ can read its value, namely
$\inity$ so there will be no branching in this case, 
and the successor node will be $\tnode_{11}$.
At $\tnode_{11}$ the only available write statement $\xvar:=3$
is selected after which $\creg:=\yvar$ (reading from $\inity$),
and $\xvar:=4$ are executed.
Notice that the trace obtained at node $\tnode_{14}$ is $\trace_1$
and the observation sequence is
$
\xvar:=2\;
\areg=0\;
\breg=0\;
\xvar:=3\;
\creg=0\;
\xvar:=4$.
The recursive call to $\explore$ 
from  $\tnode_{13}$ to  $\tnode_{14}$ 
will return immediately to $\tnode_{13}$  since there are no enabled statements 
left in the program, after which
$\declarepostponed$ will be called 
(line~\ref{explore:conflict:line} of $\explore$).
The latter will detect that $x:=4$ is postponed w.r.t.\ $\reg:=\xvar$
(line~\ref{declare:postponed:find:read:line} of $\declarepostponed$).
The fact that $x:=4$ is postponed w.r.t.\ $\reg:=\xvar$ is 
marked by the red arrow in Figure~\ref{dpor:example:fig}.
The $\declarepostponed$ procedure executes the for-loop at line~\ref{declare:postponed:main:loop:line}
and creates  a schedule $\schedule_1$ given by $\creg=0\;\xvar:=4\;\areg=4$ 
(the event $\creg=0$ precedes $\xvar:=4$ w.r.t.\ $\po$).
The schedule $\schedule_1$ is added to the set of schedules of $\areg:=\xvar$.
Similarly, when the recursive call returns to the node 
$\tnode_{11}$, the schedule $\schedule_2$ containing the sequence
$\breg=0\;\xvar:=3\;\areg=3$ will be added to the set of schedules of $\areg:=\xvar$.

When the calls of the nodes in the left-most branch 
(nodes $\tnode_{10}$--$\tnode_{14}$)
have all returned and we are at node $\tnode_{1}$ again,
$\explore$ will make a recursive call to node $\tnode_{20}$
corresponding to the second iteration of the for-loop at line~\ref{explore:aval:line}.
From $\tnode_{20}$, the execution will continue similarly
to the branch $\tnode_{10}$--$\tnode_{14}$, creating the branch
$\tnode_{20}$--$\tnode_{24}$ obtaining the trace $\trace_2$.
In particular, two schedules will be obtained namely
$\creg=0\;\xvar:=4\;\areg=4$ at $\tnode_{21}$ and
$\breg=0\;\xvar:=3\;\areg=3$ at $\tnode_{23}$.
However, these two schedules are identical (and hence trivially equivalent)
to the schedules $\schedule_1$ and $\schedule_2$ respectively, and therefore
they will not be added to the set of schedules of $\areg:=\xvar$.

When the recursive call from $\tnode_{20}$ has returned to $\tnode_1$,
the for-loop at line~\ref{explore:aval:line}  will terminate and $\explore$ moves to
line~\ref{explore:if:read:for2:line}  where
it considers the set of schedules of $\areg:=\xvar$ (in some order).
In the example, it selects  $\schedule_2$ first and calls $\rschedule$,
which will create the nodes $\tnode_{30}$, $\tnode_{31}$, and
$\tnode_{32}$.
In particular, $\rschedule$ will set  $\swapof{\areg:=\xvar}$ to false
(line~\ref{rschedule:non:swap:line}).
From $\tnode_{32}$ the call returns  to $\explore$
(line~\ref{rschedule:explore:line} of $\rschedule$) and the nodes $\tnode_{33}$
and $\tnode_{34}$ will be created leading to the trace $\trace_3$.
Although the write event $\xvar:=4$ is potentially postponed
w.r.t.\ $\areg:=\xvar$, the corresponding schedule will not
be added since the $\swap$ flag
of $\areg:=\xvar$ is false (line~\ref{declare:postponed:find:read:line} of $\declarepostponed$).
This means that 
the algorithm will not create
the dotted branch ($\tnode_{52}$--$\tnode_{54}$) in Figure~\ref{dpor:example:fig}.

When the recursive call returns from $\tnode_{30}$ to $\tnode_1$
the schedule $\schedule_1$ will be run in a similar manner to $\schedule_2$
resulting the right-most branch ($\tnode_{40}$--$\tnode_{44}$) and the trace
$\trace_4$.

\paragraph{Optimality.}
The test at line~\ref{checking:line} of $\declarepostponed$ is necessary.
It prevents adding
the schedules from $\tnode_{23}$ and $\tnode_{21}$.
These two schedules would lead to duplicates of the traces
$\trace_3$ and $\trace_4$.
Checking the status of $\swap$  is also necessary
(line~\ref{declare:postponed:find:read:line} of $\declarepostponed$).
In our example, this prevents adding the nodes $\tnode_{52}$--$\tnode_{54}$.
Adding these nodes would result in $\trace_5$ which is identical to $\trace_4$.

\subsection{Properties of the DPOR Algorithm}
\label{properties:algorithm:section}
%\patodo{Check please!}

\noindent
{\bf Soundness}
The algorithm is sound in the sense
is that if we initiate $\explore$ from a 
configuration $\conf$, the empty trace $\emptytrace$,
and the empty sequence of observations $\emptyword$, then
$\trace\in\cdenotationof{\ra}{\conf}$ 
for all explored traces $\trace$.
This follows from the fact that the exploration uses
the $\cmovesto{}$ relation
(cf.\ lines~\ref{explore:create:event:line} and~\ref{explore:aval:line}
in Algorithm~\ref{dpor:fig} and
line~\ref{rschedule:execute:line} in Algorithm~\ref{rschedule:fig}).

\medskip
\noindent
{\bf Optimality}
The algorithm is optimal in the sense that,
for any two different recursive calls to $\explore$ with arguments
$\tuple{\conf_1,\trace_1,\pth_1}$
and $\tuple{\conf_2,\trace_2,\pth_2}$, if $\conf_1$
and $\conf_2$ are terminal then $\weakof{\trace_1}\neq\weakof{\trace_2}$.
This follows from the following properties:
(i)
Each time we run the for-loop at line~\ref{explore:aval:line} in Algorithm~\ref{dpor:fig},
the read event $\event$ will read from a different 
write event.
(ii)
In each schedule $\schedule\in\scheduledof\event$ at line~\ref{explore:if:read:for2:line}  in Algorithm~\ref{dpor:fig}, the event
$\event$ reads from a write event $\event'$ that is different
from all the write events from which $\event$ reads at 
line~\ref{explore:aval:line}.
(iii)
Any two schedules added to $\scheduledof\event$ at line~\ref{declare:postponed:revent:line}
in Algorithm~\ref{declare:postpone:fig}
will have a read event reading from two different 
write events.
%\phongtodo{where is the proof of (iii) in the appendix?}
%The formal proof of completeness can be found in the appendix.

\medskip
\noindent
{\bf Deadlock-Freedom}
If $\explore$ is called with parameters
$\tuple{\conf,\trace,\pth}$ where $\conf$ is not terminal
then it will produce a subsequent recursive call.
This follows directly from Lemma~\ref{saturated:deadlock:lemma}.

\medskip
\noindent
{\bf Completeness}
The algorithm is complete in the sense that
for any input configuration $\conf$ it will produce all weak
traces corresponding to terminating runs from $\conf$.
More precisely,
for any configuration $\conf$, terminated run
$\run\in\runs(\conf)$, and total trace $\trace$ with $\trace\models\run$,
$\explore(\conf,\emptytrace,\emptyword)$
 will produce a recursive call 
$\explore(\conf',\trace',\pth)$
for some
terminal $\conf'$, $\trace'$, and $\pth$
where $\weakof{\trace'}=\weakof{\trace}$.
%
%The formal proof of completeness can be found in Section~\ref{completeness:appendix}.
%\phongtodo{How to rewrite this sentence?}

\section{Experimental Evaluation}
\label{experiment:section}

In this section,
we describe 
the implementation of 
our new techniques,
including 
the saturated semantics
in~\cref{saturated:section}
and the new optimal DPOR algorithm
w.r.t.\ weak traces
in~\cref{dpor:section},
as  a tool called 
\textsc{Tracer} 
(weak TRace ACquirE Release).
%, an extension to \textsc{CDSChecker}.
We evaluate the effectiveness of  \textsc{Tracer}  
in practice, 
by comparing its performance
with two other 
stateless model checking tools
running under the Release-Acquire semantics, 
namely 
\textsc{CDSChecker}~\cite{NoDe:toplas16} and {\textsc{Rcmc}~\cite{KLSV:popl18}.

\medskip
\noindent{\bf The \textsc{Tracer} tool}
\textsc{Tracer} can be used
to unit test portions
of concurrent code
written in the Release-Acquire fragment of the C/C++11 memory model
to discover
which behaviors the memory model allows.
By 
analyzing
the set of weak traces
generated during
the exploration
of  programs,
\textsc{Tracer} can  detect runtime errors such as user-provided assertion violations, deadlocks (with  standard definition as in~\cite{Silberschatz:2012:OSC:2490781}), and  buggy accesses  on uninitialized memory locations.
We emphasize that 
all  deadlocks
detected
by
\textsc{Tracer}
 actually occur in the provided programs
and they are
not due to our DPOR algorithm.

\textsc{Tracer} 
is constructed as  dynamically-linked shared libraries which implement the C/C++11 
\texttt{acquire} and \texttt{release} accesses types in the \texttt{<atomic>} library~\cite{ISO:2012:III}
 and portions of the other thread-support libraries of C/C++11 (e.g., \texttt{<threads>}
and \texttt{<mutex>}).
In more detail, 
it instruments any concurrency-related API calls  
such as write and read accesses
to shared atomic  variables,
 thread creations and thread joinings 
in  input programs
into function calls provided by self-implemented utility libraries.
%\bjcom{You probably mean ``utility libraries''}
%
At runtime,
\textsc{Tracer} 
determines 
the next possible transitions
and
the values returned by atomic memory operations
following the saturated semantics in~\cref{saturated:section}.
Furthermore, \textsc{Tracer} 
controls the exploration
of the input program until it
has explored the whole set of  weak traces,
using the DPOR algorithm in~\cref{dpor:section}.
%As shown by \textsc{CDSChecker},
To verify the correctness of the instrumented libraries in \textsc{Tracer}, 
we substitute
the corresponding  headers (e.g.,\ for atomic operations) in GCC and Clang
with
 the headers of the  libraries in \textsc{Tracer}.
We observe no behavioral differences in our benchmarks when
we compile and run them using GCC or Clang with and without the instrumented libraries. 
Finally, we note that
 \textsc{Tracer}
 partly uses 
 some source code
 from \textsc{CDSChecker} 
 to preprocess and handle input programs
 at the first step
 in  its model checking progress.

%\textsc{Tracer} is a stateless model checker for concurrent program which exhaustively explores the behaviours of code under the Acquire-Release semantics. 
%%
%It can used to detect assertion violations, deadlock errors, and  buggy accessing  on uninitialized memory locations. 
%

%\phongtodo{in the dealock-freedom of the algorithm, we must say that the DPOR is deadlock free if the input program is deadlock free}

\medskip
\noindent{\bf  \textsc{CDSChecker} and \textsc{Rcmc} tools}
We compare \textsc{Tracer} with two other state-of-the-art stateless model checking tools  running under the RA semantics, namely \textsc{CDSChecker}~\cite{NoDe:toplas16} and \textsc{Rcmc}~\cite{KLSV:popl18}.
% 
%Similarly to \textsc{Tracer}, CDSChecker maintains partial coherence orders, but not in an
%optimal way. It can generate $\ra$-inconsistent executions, which must afterwards
%be validated.
%As stated above,
%\textsc{CDSChecker}
%might generate 
%a lot of redundant executions
%than \textsc{Tracer} (see~\cref{litmus:section}-\cref{industry-program:section}).
%
\textsc{CDSChecker}  implements
a DPOR algorithm
that supports concurrent programs  running under the RA semantics,
but, as illustrated in~\cref{litmus:section}--\cref{industry-program:section},
might generate a lot of redundant executions
in a less efficient way than \textsc{Tracer}.
\textsc{Rcmc} has two options: \textsc{Rc11} and \textsc{Wrc11}.
The \textsc{Rc11} option 
 generates only consistent executions (w.r.t. RA) by maintaining total coherence orders.
Meanwhile, the \textsc{Wrc11} option maintains partial coherence orders, which may
create inconsistent executions which are then not validated (see~\cref{litmus:section}).
\textsc{Rcmc}/\textsc{Rc11} is optimal in the sense that, in the
absence of RMWs, they explore each
consistent  Shasha-Snir trace~\cite{ShSn:parallel} exactly once.
Observe that \textsc{Rcmc} has limited  support for thread creation and joining~\cite{rcmc:dicuss:private}. Thus, 
we do some engineering transformations to simulate these operations.

\medskip
\noindent{\bf Experiment setup}
We compare
\textsc{Tracer} with
  \textsc{CDSChecker} and {\textsc{Rcmc} 
on six different categories
of benchmarks as follows.
\begin{enumerate}
\item
 In~\cref{litmus:section}, we  apply  \textsc{Tracer},   \textsc{CDSChecker}, and \textsc{Rcmc}   
 to a set  consisting of thousands of litmus tests (i.e., small programs) taken from~\cite{alglaveMT14}.
These experiments
are mainly used for two purposes: (i) to provide sanity checks of the correctness of the used tools, and (ii) to compare the performance  of these tools on small programs running under  RA. 
\item
In~\cref{classic:section},
we run
the  tools   on
two standard programs taken from \cite{FG:dpor}. These  benchmarks have been widely used to compare  different stateless model-checking tools  (e.g.~\cite{SKH:acsd12,abdulla2014optimal,KLSV:popl18}).
\item
In~\cref{larger:tests:section}, we run %\textsc{Tracer}, \textsc{CDSChecker} and \textsc{Rcmc} 
the tools  on 
 a collection of concurrent benchmarks taken from the TACAS competition  on Software Verification~\cite{svcomp:18}.
These benchmarks are C/C++ programs  with 50-100 lines of code 
and  used  
by many tools (e.g.~\cite{DBLP:conf/oopsla/Huang016,DBLP:conf/cav/AbdullaAJL16,AlglaveKT13,tacas15:tso}).
\item
In~\cref{synthetic:bench:section}, 
we evaluate
the  tools
on two synthetic programs (one of them is taken  from~\cite{NoDe:toplas16}). We use these benchmarks to show the benefits  of using weak traces  used by \textsc{Tracer} (see~\cref{model:section})  instead of 
Shasha-Snir traces~\cite{ShSn:parallel} used by \textsc{Rcmc}.

\item
In~\cref{more:synthetic:bench:section},
we apply 
%\textsc{Tracer}, \textsc{CDSChecker} and \textsc{Rcmc} 
the tools
to three parameterized programs 
to study their behaviors
when 
increasing the numbers of 
read and write operations
and 
 threads
in 
the programs.   
%expose more differences between the three tools.
\item
In~\cref{industry-program:section}, we use the  tools
to analyze
more extensive and more challenging benchmarks, containing several implementations of  concurrent data structures.
Some of them were used in the evaluation of   \textsc{CDSChecker}~\cite{NoDe:toplas16,DBLP:conf/oopsla/DemskyL15}  and \textsc{Rcmc}~\cite{KLSV:popl18}.
The other benchmarks are high-performance  starvation-free algorithms in~\cite{c11:bench,tidex}.  
\end{enumerate}

We conduct  all experiments on
 a Debian 4.9.30-2+deb9u2 machine 
with an Intel(R) Core(TM) i7-3720QM CPU (2.60 GHz) and 16 GB of RAM. 
\textsc{Tracer} and \textsc{CDSChecker}
have been compiled  with GCC  v6.3.0.
We use the artifact-evaluated version of \textsc{Rcmc}~\cite{KLSV:popl18} 
that
is compiled
with Clang and LLVM v3.8.1.
We use the
 argument \texttt{-u 0} in all benchmarks
 for both \textsc{Tracer} and \textsc{CDSChecker}
 that provides $0$
 for \texttt{release} atomic loads
 from uninitialized memory locations. 
All these tools were run on deterministic  programs with bounded executions. This is achieved  by unrolling loops in any tested program up to a specific bound. Therefore, we do not need \texttt{-m} and \texttt{-y}  arguments  to control
memory liveness and thread yield-based fairness
for \textsc{Tracer} and \textsc{CDSChecker}
 as in~\cite{NoDe:toplas16}
 and \texttt{-unroll} argument for \textsc{Rcmc}.
Finally, all experiments have been set up with a  1-hour timeout.

\subsection{Evaluation using Litmus Tests}
\label{litmus:section}

 We apply \textsc{Tracer}, \textsc{CDSChecker}, and \textsc{Rcmc}
 on a set of litmus tests, taken from~\cite{alglaveMT14}.
 Litmus tests are standard benchmarks used by many tools running on weak memory models (e.g.~\cite{alglaveMT14,DBLP:conf/pldi/SarkarSAMW11,KLSV:popl18,DBLP:conf/cav/AbdullaAJL16}).
 Typically, each litmus test contains at most $4$ threads with less than $5$ shared variables.
 In a litmus test, threads concurrently execute small sets of instructions. 
 After all threads have finished their executions, an assertion is validated 
 to check whether 
 the set of some behaviors presented by the assertion is allowed.
 Since litmus tests are small  but  contain all kinds of dependencies such as address dependency, data dependency,
 and control dependency (see~\cite{alglaveMT14}), 
they are usually used to provide sanity checks of the correctness of the tools and to compare the performance of these tools on small benchmarks. 
Observe that all the above dependencies in litmus tests can be handled under the RA semantics 
 by executing all events  following the program order and using the read-from relations.
We 
translate  litmus tests into C/C++11 by considering all writes and reads as  \texttt{release} and \texttt{acquire} atomic accesses, and
all synchronizations as  RMW accesses 
(implemented by \texttt{atomic\_fetch\_add} API calls)
to a particular fence variable.
For an unsafe test, we 
stop all tools as soon as they find the first bug since \textsc{Rcmc} does so.

\begin{table*}[t]
\centering
\small{
\begin{tabular}{  l  r c r  r r  r r r r r}
\hline
\multirow{2}{*}{{\bf Tool}}  &\;\;\;\;\;& \multirow{2}{*}{{\bf Wrong answer}} &\;\;\;\;\;& \multirow{2}{*}{{\bf Execs} }  &\;\;\;\;\;\;& \multirow{2}{*}{ {\bf Time}}   &\;\;\;\;\;& {\bf Average Execs} 	&\;\;\;\;\;& {\bf Average Time}	 \\ 
					&&		&&		&&	 &\;\;& $(\textit{\it Execs}/\textit{Nums of tests})$ 	&\;\;& $(\textit{Time}/\textit{Execs})$	 \\\hline 
\textsc{CDSChecker} 	&&  0 			&& 2 383 704 		&& 6m28s 		&&	295.59 		&&	{\bf 0.16ms} \\
\textsc{Rcmc/Rc11} 		&& 0 			&& 727 942		&&  14m00s	  	&&	90.27 		&&	1.15ms	\\
\textsc{Rcmc/Wrc11} 	&& 81			&& 668 574 		&& 12m15s 		&& 	82.91		&& 	1.10ms	\\
\textsc{Tracer}  			&&  0 			&& {\bf 521 288}  	&& {\bf 3m03s} 		&&	{\bf 64.64}		&&	0.35ms	\\ 
\hline 
\end{tabular}
}
\caption{Comparison of the three tools when running on  the litmus tests. The column \textit{Wrong answer}   corresponds to  the number of wrong answers.
The {\it Execs} (resp.\ {\it  Average Execs})  column 
is the total number of explored executions
(resp.\ the average explored executions per  test). 
The  {\it Time}  (resp.\ {\it  Average Time}) column
is the total  time  (resp. the average  time per execution) 
for running the tests without accumulating the compilation time.
For \textsc{Rcmc}, these running time are approximate because we cannot measure only the running time
and exclude the compilation time.
We believe that
the compilation time of \textsc{Rcmc} is far smaller than the running time.
The best number of executions and running time of the tools  are given in bold font.}
\label{litmus:experiment}
\end{table*}

Table~\ref{litmus:experiment} summarizes the results of different tools on 8064  litmus tests.
First of all, 
%as depicted in Table~\ref{litmus:experiment},
all tools, except \textsc{Rcmc/Wrc11}, are consistent w.r.t.\ the RA semantics.
\textsc{Rcmc/Wrc11} is inconsistent, generating  
81 wrong answers\footnote{Based on our reported results, newer versions of \textsc{Rcmc}
can fix some of the 81 wrong answers~\cite{rcmc:dicuss:private}.}.
A {\it wrong answer} denotes  that the outcome of the tool in a test is different 
from the  expected outcome\footnote{We generated the expected outcomes using the Herd tool together with the set of  RA-axioms provided  in~\cite{DBLP:conf/popl/LahavGV16}. Since Herd is not a stateless model checker and it is significantly
slower in these tests, we do not include Herd in our comparison.}.
In the above 81 wrong answers, \textsc{Rcmc/Wrc11} reports unsafe results for the corresponding safe tests. 
The  inconsistency of \textsc{Rcmc/Wrc11} is due to the imprecise handling of coherence
orders.
In fact, \textsc{Rcmc/Wrc11} sometimes does not generate  coherence orders
that are needed to preserve  RA consistency, e.g.,\ the coherence edges in the trace $\trace_7$ in Figure~\ref{saturated:semantics:fig}.
The missing coherence orders help \textsc{Rcmc/Wrc11} to reduce the number of executions, 
but can also lead to inconsistent executions.
Indeed,  \textsc{Rcmc/Wrc11} allows the inconsistent trace $\trace_9$ in Figure~\ref{saturated:semantics:fig}
for the program given in Figure~\ref{overview:prog:fig}.
Secondly, \textsc{Tracer} explores fewer executions and 
has a better performance  than the other tools.
For example, it generates 22\% fewer executions  than \textsc{Rcmc/Wrc11}
(that can be observed based on the columns \textit{Execs} and/or \textit{Average Execs} in Table~\ref{litmus:experiment})
which in turn is 
exploring fewer executions than the remaining tools. 
Moreover,
\textsc{Tracer} is also 4 times faster than \textsc{Rcmc/Wrc11}.  
Thirdly, 
we compare three tools 
on
{\it the average running time
spent by each tool on the exploration of an execution} (the {\it Average Time} column  in Table~\ref{litmus:experiment}).
The
average running time
per  execution
is calculated 
using the formula
$${\it Average Time}:={\it Time}/{\it Execs}$$
i.e.\ it is the average time a tool needs to explore one execution.
We observe that
based on  ${\it Average Time}$,
 \textsc{Tracer} 
is
about 3 times faster than \textsc{Rcmc}
but slower than \textsc{CDSChecker}.
The reason is that
 \textsc{CDSChecker}  can explore  
  prefixes of some redundant executions before ending them and therefore  \textsc{CDSChecker} spends very little  time on these small incomplete executions.  
 Meanwhile,
 all explored executions by \textsc{Tracer} and \textsc{Rcmc} are complete  since all litmus tests only have assertions at the end of programs.

Since \textsc{Rcmc/Wrc11} is inconsistent w.r.t.\  the RA semantics (cf.\ Table~\ref{litmus:experiment}), 
we exclude it from further comparisons.
For the sake of convenience, in the following, we use \textsc{Rcmc} for \textsc{Rcmc/Rc11}.

%%%%%%%%%%%%%%%%%%%%%%%%%%%%%%%%%%%%%%%%%%%%%%%%%%%%
%%%%%%%%%%%%%%%%%%%%%%%%%%%%%%%%%%%%%%%%%%%%%%%%%%%%

\begin{table*}[tb]
\centering
\small
\begin{tabular}{  rl   rr  rrrrrrr  rrrrrrr r}
\hline
& \multirow{2}{*}{{\bf Program}} &\;\;\;\;\;\;\;\;& \multirow{2}{*}{{\bf LB}}   	&\;\;\;\;& \multicolumn{5}{c}{{\bf Executions Explored}} &\;\;\;\;\;\;& \multicolumn{5}{c}{{\bf Total Running Time}} 		&\\ 
\cline{6-10}
\cline{12-16}
&					&& 						&& \textsc{CDSChecker} &\;\;\;\;& \textsc{Rcmc} &\;\;\;\;& \textsc{Tracer}		&& \textsc{CDSChecker} &\;\;\;\;& \textsc{Rcmc} &\;\;\;\;& \textsc{Tracer} 			&\\ 
\hline %
& Indexer(13)			&& 20					&& 190 && {\bf 64} && {\bf 64}					&& 0.15s && 0.62s && {\bf 0.04s}					&\\
& Indexer(14)			&& 20					&& 3 075 && {\bf 512} && {\bf 512}				&& 2.72s && 11.68s && {\bf 0.38s}					&\\
& Indexer(15)			&& 20					&& 48 261 && {\bf 4 096} && {\bf 4 096} 			&& 45.32s && 4m22s && {\bf 2.98s}					&\\
& Indexer(16)			&& 20					&& 740 889 &&   {\it t/o} &&  {\bf 32 768}			&& 12m21s &&{\it t/o} && {\bf 25.78s}				&\\
\hline
& Filesystem(16)		&& 20					&& 27 && {\bf 8} && {\bf 8}  					&& 0.04s && 0.05s && {\bf 0.01s}					&\\ 
& Filesystem(19)		&& 20					&& 728 && {\bf 64} && {\bf 64}					&& 0.52s && 2.02s &&{\bf 0.03s}					&\\ 
& Filesystem(22)		&& 20					&& 19 678 && {\bf 512} && {\bf 512}  			&& 13.53s && 2m29s && {\bf 0.33s}					&\\ 
& Filesystem(25)		&& 20					&& 531 415 && {\it t/o} && {\bf 4096}  			&& 7m04s && {\it t/o} && {\bf 3.12s}					&\\ 
\hline
\end{tabular}
\caption{
Comparison of the performance of the three tools  when running on two standard benchmarks: \texttt{Indexer(N)} ~\cite{svcomp:18} and \texttt{Filesystem(N)} 
~\cite{FG:dpor}.
The {\it LB} column indicates how many times a loop is unrolled.  
If a tool runs out of time, we put  {\it t/o}.
The best number of executions and running time  for each benchmark are given in bold font.
}
\label{classic-benchs:table}
%\vspace{-6mm}
\end{table*}

\subsection{Evaluation using two Standard Benchmarks}
\label{classic:section}

%Table~\ref{classic-benchs:table}
%gives
 %the results
We compare \textsc{Tracer} with  \textsc{CDSChecker} and \textsc{Rcmc} 
on 
two standard benchmarks, namely
\texttt{Indexer(N)}\footnote{
We do experiments with the \texttt{Indexer(N)} benchmark from~\citet{svcomp:18} since it is similar to the original one and widely used in the verification community.
The \texttt{Indexer(N)}   in~\cite{KLSV:popl18} implements a different version.}
and
\texttt{Filesystem(N)},
introduced in~\cite{FG:dpor}.
These  benchmarks  have been widely used 
to compare different DPOR techniques
under the SC semantics, e.g.\ \cite{FG:dpor,SKH:acsd12,abdulla2014optimal}. 
 %
 %While there are several threads searching a free slot to allocate a data block
%in filesystem,
%there are several threads searching and inserting messages  into a hash table in indexer.
We parameterize these benchmarks by the number of threads.
For \texttt{Indexer(N)}, we use 13, 14, 15, and 16 threads.
%\footnote{These numbers were chosen to show the difference between the different tools}.
%
For \texttt{Filesystem(N)},
 we
use 16, 19, 22, and 25 threads. 
The reason of the parameterization
is that more threads  generate more 
conflicts between their write and read operations, 
and therefore more non-equivalent executions are needed to be explored by different tools. 
For example, there will be no conflicts
between the operations of threads
in \texttt{Indexer(N)} with  11 threads
and in  \texttt{Filesystem(N)} with  13 threads.

In Table~\ref{classic-benchs:table}, we report
the number of executions that the three tools explore as well as the
time needed  by the tools to explore them. 
We observe that \textsc{Tracer}  always examines  optimal numbers of executions w.r.t.\ weak traces
and has a better performance than \textsc{CDSChecker} and \textsc{Rcmc} in all benchmarks.
The difference between \textsc{Tracer} and the other tools  becomes more evident when we increase the number of threads.
In both \texttt{Indexer(N)} and  \texttt{Filesystem(N)}, \textsc{Rcmc} can discover the same numbers of executions as \textsc{Tracer};
however, its total running time and the average time per execution
are two orders of magnitude slower than \textsc{Tracer} on average.
Observe that the used version of  
\texttt{Indexer(N)} program 
contains lock primitives (as it is the case in~\citet{svcomp:18}). This version is different from the one used in~\cite{KLSV:popl18} and this explains the difference  in the  obtained performance for   \textsc{Rcmc} between  the one reported in Table~\ref{classic-benchs:table} and the one published in~\cite{KLSV:popl18}.  
%We conjuncture that this is due to the  way \textsc{Rcmc}  handles lock primitives. 
Based on our discussion with \textsc{Rcmc}'s authors~\cite{rcmc:dicuss:private}, we conjecture that this performance issue  is due to the way
\textsc{Rcmc}  handles lock primitives.
Finally, 
\textsc{CDSChecker} explores more executions than \textsc{Tracer} in all benchmarks,
and it is about 20 times slower than \textsc{Tracer}.

%%%%%%%%%%%%%%%%%%%%%%%%%%%%%%%%%%%%%%%%%%%%%%%%%%%%
%%%%%%%%%%%%%%%%%%%%%%%%%%%%%%%%%%%%%%%%%%%%%%%%%%%%

\subsection{Evaluating using SV-COMP Benchmarks}
\label{larger:tests:section}

\begin{table*}[tb]
\centering
\small
\begin{tabular}{  rl   rr  rrrrrrr  rrrrrrr r}
\hline
& \multirow{2}{*}{{\bf Program}} &\;\;\;\;\;& \multirow{2}{*}{{\bf LB}}   	&\;\;\;& \multicolumn{5}{c}{{\bf Executions Explored}} &\;\;\;& \multicolumn{5}{c}{{\bf Total Running Time}} 				&\\ 
\cline{6-10}
\cline{12-16}
&					&& 						&& \textsc{CDSChecker} &\;\;\;& \textsc{Rcmc} &\;\;\;& \textsc{Tracer}			&& \textsc{CDSChecker} &\;\;\;& \textsc{Rcmc} &\;\;\;& \textsc{Tracer} 		&\\ 
\hline
& Pthread\_demo		&& 10					&& 184 758 && {\bf 184 756} && {\bf 184 756}				&& 24.96s && 1m03s && {\bf 18.00s}				&\\
& Gcd				&& 8						&&  8 814 044&& {\bf 1 162 333} && {\bf 1 162 333}			&&  38m20s && 7m07s && {\bf 2m48s}				&\\
& Fibonacci			&& 6						&& {\it t/o} && {\bf 525 630} && {\bf 525 630}				&& {\it t/o} && {\bf 31.06s} && 57.95s					&\\
& Szymanski			&& 6						&& {\it t/o} && 26 037 490 && {\bf 12 209 410}				&& {\it t/o} && 44m52s && {\bf 14m54s}				&\\
& Dekker				&& 10					&& 7 306 447 && {\bf 3 121 870} && {\bf 3 121 870}  		&& 15m25s && 5m12s && {\bf 4m52s}				&\\
& Lamport				&& 8						&& {\it t/o} &&  6 580 870 && {\bf 3 372 868}				&& {\it t/o} && 14m40s && {\bf 6m58s}				&\\
& Sigma(5)			&& 5						&& 1 279 && 945 &&	{\bf 120}							&& 0.09s && 0.16s &&  {\bf 0.01s}					&\\
& Peterson			&& 6						&& {\it t/o} && {\bf 1 897 228} &&  {\bf 1 897 228} 			&& {\it t/o} && 3m16s && {\bf 3m15s}					&\\
& Stack\_true			&& 12					&& 2 704 157 && {\bf 2 704 156} && {\bf 2 704 156} 			&& 19m03s && 54m25s && {\bf 10m12s}				&\\
& Queue\_ok			&& 12					&& 581 790 && {\it t/o} && {\bf 362 880}					&& 33m27s && {\it t/o} && {\bf 12m52s}				&\\
\hline
\end{tabular}
\caption{
Comparison of the performance of the three tools  when running on the TACAS Competition on
Software Verification (SV-COMP)  benchmarks. 
The {\it LB} column 
and the {\it t/o} entry have the same meanings as in Table~\ref{classic-benchs:table}. 
}
\label{tascas-benchs:table}
%\vspace{-6mm}
\end{table*}

We compare \textsc{Tracer} with  \textsc{CDSChecker} and \textsc{Rcmc} on 
the  set of  concurrent benchmarks from~\citet{svcomp:18} (the TACAS Software Verification competition 2018), cf. Table~\ref{tascas-benchs:table}.
These benchmarks consist of  ten  programs 
that
are written in C/C++
with
50-100 lines of code 
and  used  
by many tools (e.g.~\cite{DBLP:conf/oopsla/Huang016,DBLP:conf/cav/AbdullaAJL16,AlglaveKT13,tacas15:tso}).
The primary challenge in these benchmarks is to handle   a large number of  executions
 that  are needed to be explored.
For example, \textsc{Rcmc} generates 26 million  executions for the \texttt{Szymanski} benchmark.
As stated in~\cref{litmus:section}, 
for  unsafe benchmarks,
\textsc{Rcmc}
 stops the exploration
as soon as it detects the first assertion violation.
To fairly compare the efficiency of different DPOR approaches, as it was done in~\cite{KLSV:popl18,abdulla2014optimal}, 
we remove all the assertions in the benchmarks
to
let all tools {\it exhaustively} explore all possible executions.

From Table~\ref{tascas-benchs:table}, we can see that \textsc{Tracer} explores  smaller numbers of executions
than the two other tools in all cases.
As a consequence, \textsc{Tracer} has
the best performance in 9 of 10 examples.
Meanwhile, \textsc{CDSChecker} times out in 4 of 10 cases.
The main reason for these timeouts is that \textsc{CDSChecker} needs to explore a large number of executions.
This can be seen in the  \texttt{Gcd} example where the number of executions generated
by \textsc{CDSChecker} is huge, compared to those produced by  \textsc{Rcmc} and \textsc{Tracer}. 
On two benchmarks,
 \texttt{Pthread\_demo} and
 \texttt{Stack\_true},
 where the three tools generate almost the same numbers of executions,
\textsc{CDSChecker} is faster than \textsc{Rcmc} but still less efficient than \textsc{Tracer}.
The only benchmark in which
\textsc{Tracer} does not have the best performance
is  \texttt{Fibonacci}.
In this benchmark, the numbers of  executions explored
by  \textsc{Tracer} and \textsc{Rcmc}
are exactly the same,
and \textsc{Rcmc} is slightly faster.
However, in general 
\textsc{Tracer} generates 40\% fewer number of executions and it is about 3 times faster than \textsc{Rcmc}.  
Interestingly,
\textsc{Tracer} is 2 times faster than \textsc{Rcmc} in  the average time per execution,
which  coincides with the litmus tests in~\cref{litmus:section}.
 
% 25178371,  2353.25 \textsc{Tracer} 10699.4
% 42215278, 7866.22  5366 \textsc{Rcmc}

 \subsection{Evaluation using Synthetic Benchmarks}
 \label{synthetic:bench:section}

Next, we expose more differences between the three tools on two
synthetic benchmarks given in Figure~\ref{synthetic:code}. The first one is   \texttt{N\_writers\_a\_reader(N)}
benchmark, taken from~\cite{NoDe:toplas16}. The results, for 7, 8, 9, and 10 threads, are given in Table~\ref{synthetic-benchs:table}. 
Since this benchmark does not contain any loops, we do
not use loop unrolling and therefore
 do not
 show the LB column as in Tables~\ref{classic-benchs:table} and~\ref{tascas-benchs:table}. 
Since \textsc{Rcmc}  is optimal w.r.t.\ Shasha-Snir traces in the absence of RMWs,
 the number of  executions explored by \textsc{Rcmc} is exactly (factorial) $(N+1)!$ here.
 This number corresponds to the number of possible combinations of
 all feasible $\po$, $\rf$, and total $\co$.
 However, one can easily see that there are only $N+1$ possible values
 for the read,
 presented by $\avar=0$, $\avar=1$, $\cdots$, $\avar=N$.
 In fact, 
 the total coherence order is irrelevant in this benchmark and only the read-from relation is important.
 Therefore,
 in contrast to \textsc{Rcmc}, \textsc{Tracer}, which is optimal w.r.t.\ weak traces, precisely explores $N+1$ executions   (i.e.,  linear).

  \begin{figure}[tb]
\begin{minipage}[b]{0.38\linewidth}
\center
\begin{lstlisting}[style=CStyle]
/* initially: x=0 */
atomic_int x;

/* N writers*/
void writers(void *arg) 
{ 
	/* tid is from 1 to N */ 
	int tid = *((int *)arg);
	x.store(tid, release); 
}

void reader(void *arg) 
{ 
	int a = x.load(acquire);
}

\end{lstlisting}

\end{minipage}
\hspace{8mm}
\begin{minipage}[b]{0.38\linewidth}
\center
\begin{lstlisting}[style=CStyle]
/* initially: x=0 */
atomic_int x; 

/* 2 writers */
void writers(void *arg) 
{ 
	for (int i=0; i<N; i++)
		  x.store(1, release); 
}

void reader(void *arg) 
{ 
	int a = x.load(acquire);
	int b = x.load(acquire);
}


\end{lstlisting}

\end{minipage}
\caption{C/C++11 codes of \texttt{N\_writers\_a\_reader(N)} (left) and \texttt{Redundant\_co(N)} (right).}
\label{synthetic:code}
\end{figure}

 \begin{table*}[tb]
\centering
\small
\begin{tabular}{  rl     rrrrrrr  rrrrrrr r}
\hline
& \multirow{2}{*}{{\bf Program}}  		&\;\;\;\;\;\;& \multicolumn{5}{c}{{\bf Executions Explored}} 		&\;\;\;\;\;\;& \multicolumn{5}{c}{{\bf Total Running Time}} 				&\\ 
\cline{4-8}
\cline{10-14}
& && \textsc{CDSChecker} &\;\;\;& \textsc{Rcmc} &\;\;\;& \textsc{Tracer}			&& \textsc{CDSChecker} &\;\;\;& \textsc{Rcmc} &\;\;\;& \textsc{Tracer} 		&\\ 
\hline
& N\_writers\_a\_reader(7)			&&  {\bf 8} && 40 320 && {\bf 8}  						&& 0.01s && 0.46s && {\bf 0.00s}					&\\
& N\_writers\_a\_reader(8)			&&  {\bf 9} && 362 880 && {\bf 9}  						&& 0.01s && 4.19s && {\bf 0.00s}					&\\
& N\_writers\_a\_reader(9)			&&  {\bf 10} && 3 628 800 && {\bf 10}  					&& 0.01s && 46.13s && {\bf 0.00s}					&\\
& N\_writers\_a\_reader(10)			&&  {\bf 11} && 39 916 800 && {\bf 11}  					&& 0.01s && 9m35s && {\bf 0.00s}					&\\
\hline
& Redundant\_co(5)					&&  581 && 16 632 &&  {\bf 91}							&& 0.03s  && 0.39s && {\bf 0.01s}					&\\
& Redundant\_co(10)				&& 10 631  && 42 678 636 && {\bf 331}  					&& 0.64s && 23m56s && {\bf 0.02s}					&\\
& Redundant\_co(15)				&&  59 056 && {\it t/o} && {\bf 721}  						&& 4.57s && {\it t/o} && {\bf 0.06s}					&\\
& Redundant\_co(20)				&& 197 231 && {\it t/o} && {\bf 1 261}  					&& 20.27s && {\it t/o} && {\bf 0.14s}					&\\
\hline

\end{tabular}
\caption{
Comparison of the performance of the three tools  when running on  two synthetic  programs.
%:  
%\texttt{N\_writers\_a\_reader(N)} and \texttt{Redundant\_co(N)}.
The {\it t/o} entry has the same meaning as in Table~\ref{classic-benchs:table}.
}
\label{synthetic-benchs:table}
%\vspace{-6mm}
\end{table*}
 
The second benchmark is called \texttt{Redundant\_co(N)}  and its code is given in Figure~\ref{synthetic:code}. We use this example to emphasize more  the benefit of using  weak traces.
As depicted in Table~\ref{synthetic-benchs:table},
while the number of Shasha-Snir traces and  the number of explored traces by \textsc{Rcmc} is  $O(N!)$ (i.e., factorial),
\textsc{Tracer}  explores  only  $O(N^2)$ executions\footnote{The exact  number of weak traces in  \texttt{Redundant\_co(N)}  is $3N^2+3N+1$.}.
For instance for \texttt{Redundant\_co(10)},
\textsc{Tracer} finishes the exploration of the program in less than 1 second,
while  
\textsc{Rcmc} 
examines two orders of magnitude more 
executions and needs almost  half an hour to complete.
Since \textsc{CDSChecker} also maintains a partial coherence order as \textsc{Tracer} but not in an optimal way,
\textsc{CDSChecker} can explore the same  number of executions for \texttt{N\_writers\_a\_reader(N)}  as \textsc{Tracer} but not in  \texttt{Redundant\_co(N)}.

 \subsection{Evaluation using  Parameterized Benchmarks}
 \label{more:synthetic:bench:section}

 \begin{table*}[t]
\centering
\small
\begin{tabular}{  rl   rr  rrrrrrr  rrrrrrr r}
\hline
& \multirow{2}{*}{{\bf Program}} &\;\;\;& \multirow{2}{*}{{\bf LB}}   	&\;\;\;& \multicolumn{5}{c}{{\bf Executions Explored}} &\;\;\;\;& \multicolumn{5}{c}{{\bf Total Running Time}} 		&\\ 
\cline{6-10}
\cline{12-16}
&					&& 			&& \textsc{CDSChecker} &\;\;\;& \textsc{Rcmc} &\;\;\;& \textsc{Tracer}		&& \textsc{CDSChecker} &\;\;\;& \textsc{Rcmc} &\;\;\;& \textsc{Tracer} 			&\\ 
\hline
& Sigma(6)			&&6			&& 25 357  && 10 395 && {\bf 720}						&& 2.20s && 1.96s &&  {\bf 0.04s}					&\\
& Sigma(7)			&&7			&&  605 714 && 135 135 && {\bf 5 040}					&& 1m02s  && 29.06s &&  {\bf 0.40s}				&\\
& Sigma(8)			&&8			&& 16 667 637 && 2 027 025 && {\bf 40 320}				&& 33m18s && 8m02s &&  {\bf 3.28s}				&\\
& Sigma(9)			&&9			&& {\it t/o} && {\it t/o} && {\bf 362 880}					&& {\it t/o} &&  {\it t/o} &&  {\bf 33.59s}				&\\
\hline
& Control\_flow(6)		&&0			&& 896  && 55 440 && {\bf 77}							&& 0.09s && 1.97s && {\bf 0.01s}					&\\
& Control\_flow(8)		&&0			&& 4 608  && 11 007 360 && {\bf 273} 					&& 0.53s && 7m58s && {\bf 0.03s}					&\\
& Control\_flow(10)		&&0			&&  22 528 &&{\it t/o}  && {\bf 1 045} 					&& 3.27s &&{\it t/o} && {\bf 0.16s}					&\\
& Control\_flow(12)		&&0			&&  106 496 &&{\it t/o}  && {\bf 4 121} 					&& 19.10s &&{\it t/o} && {\bf 0.79s}					&\\
\hline
& Exponential\_bug(6)	&&6			&&  983 386 && 1 203 446 && {\bf 15 601}  				&& 1m18s && 56.75s && {\bf 0.96s}					&\\
& Exponential\_bug(7)	&&7			&&   2 250 290 && 2 833 112 &&{\bf  22 841}  				&& 3m13s && 2m26s && {\bf 1.46s}					&\\
& Exponential\_bug(8)	&&8			&&  4 844 378 && 6 158 718 && {\bf 32 313} 				&& 7m15s && 5m28s && {\bf 2.23s}					&\\
& Exponential\_bug(9)	&&9			&&  9 896 954 && 12 526 576 && {\bf 44 428}  				&& 15m29s && 11m48s && {\bf 3.22s}				&\\
\hline

\end{tabular}
\caption{
Comparison of the performance of the three tools  when running on  three  parameterized benchmarks.
The {\it LB} column and the {\it t/o} entry have the same meanings as in Table~\ref{classic-benchs:table}. 
}
\label{more:synthetic-benchs:table}
%\vspace{-3mm}
\end{table*}

Table~\ref{more:synthetic-benchs:table} 
reports more experimental results of \textsc{Tracer}, \textsc{CDSChecker}, and \textsc{Rcmc}
on three parameterized benchmarks: \texttt{Sigma(N)}  (presented in Table~\ref{tascas-benchs:table}), \texttt{Control\_flow(N)} (used in~\cite{abdulla2014optimal}),
and \texttt{Exponential\_bug(N)} (presented in Figure 2 of~\cite{DBLP:conf/pldi/Huang15}).
In \texttt{Sigma(N)} and  \texttt{Control\_flow(N)}, $N$ is the number of threads used in these  benchmarks.
Meanwhile, in \texttt{Exponential\_bug(N)},
$N$ is the number of times a thread writes to  a specific variable.
In a similar way to Tables~\ref{classic-benchs:table}--\ref{synthetic-benchs:table}, Table~\ref{more:synthetic-benchs:table}   
shows that
 \textsc{Tracer} always
 explores
 smaller numbers of executions than \textsc{CDSChecker} and \textsc{Rcmc}
 in all benchmarks,
 and it has the best performance. 
The differences between  \textsc{Tracer} and the other tools becomes more clear when we increase the number of threads.
%explores the good stability in the algorithms that \textsc{Tracer} employs over \textsc{CDSChecker} and \textsc{Rcmc}.
For example,
in \texttt{Sigma(N)} benchmark,
the number of  executions explored by \textsc{Tracer}
increases only  8 times at each step, while   it increases by
 23 and 15 times for \textsc{CDSChecker} and \textsc{Rcmc} respectively.
As a consequence, 
when there are more than 7 threads,
the verification tasks of \texttt{Sigma(N)} are very challenging
for \textsc{CDSChecker} and \textsc{Rcmc} 
but can still be handled by \textsc{Tracer}
in less than 1 minute.

\subsection{Evaluation using Concurrent Data Structure  Benchmarks}
\label{industry-program:section}
 
We apply \textsc{Tracer} to  fifteen  concurrent data structure algorithms,  namely 
 \texttt{Correia\_Ramalhete(N)}, \texttt{Correia\_Ramalhete\_turn(N)}, 
\texttt{Tidex(N)}, \texttt{Tidex\_nps(N)},
\texttt{CLH\_c11(N)}, \texttt{CLH\_rwlock\_c11(N)}, \texttt{MPSC\_c11(N)},
\texttt{Ticket\_mutex\_c11(N)},
\texttt{Linux\_lock(N)}, \texttt{Barrier(N)}, \texttt{Seqlock(N)},  
\texttt{MPMC\_queue(N)}, \texttt{MCS\_lock(N)},
\texttt{Cliffc\_hashtable(N)}, \texttt{Concurrent\_hashmap(N)}
\texttt{Chase\_Lev\_dequeue(N)}, and \texttt{SPSC\_queue(N)}.
The first four algorithms are high-performance starvation-free mutual exclusion locks in~\cite{c11:bench,tidex}.
Meanwhile, \texttt{CLH\_c11(N)}, \texttt{CLH\_rwlock\_c11(N)}, and \texttt{MPSC\_c11(N)}
are C/C++11 implementations of CLH queue locks~\cite{MagnussonLH94} 
and \texttt{Ticket\_mutex\_c11(N)} is a C/C++11 implementation of Ticket Lock~\cite{Mellor-CrummeyS91}, 
taken from~\cite{c11:bench}. 
Other algorithms were used in the previous evaluations of \textsc{CDSChecker}~\cite{NoDe:toplas16,DBLP:conf/oopsla/DemskyL15} and \textsc{Rcmc}~\cite{KLSV:popl18}.
Since these  data structures can have  huge state-spaces,
 we limit the number $N$ of used threads in these algorithms to four.  We also use loop unrolling when it is necessary.
Moreover, to run all algorithms under the RA semantics,  as it was done in~\cref{litmus:section}--\cref{more:synthetic:bench:section}, we  consider 
all write and read operations as 
 \texttt{acquire} and \texttt{release} atomic  accesses.  
%
%The \texttt{m\&s\_queue(N)} is kept  parametric on the number of threads as from~\citet{NoDe:toplas16}.
%

\begin{table*}[tb]
\centering
\small
\begin{tabular}{  rl   rr  rrrrrrr  rrrrrrr r}
\hline
& \multirow{2}{*}{{\bf Program}} &\;\;\;\;\;& \multirow{2}{*}{{\bf LB}}   	&\;\;& \multicolumn{5}{c}{{\bf Executions Explored}} &\;\;\;& \multicolumn{5}{c}{{\bf Total Running Time}} 		&\\ 
\cline{6-10}
\cline{12-16}
&		&& 				&& \textsc{CDSChecker} &\;\;\;& \textsc{Rcmc} &\;\;\;& \textsc{Tracer}	&& \textsc{CDSChecker} &\;\;\;& \textsc{Rcmc} &\;\;\;& \textsc{Tracer} &\\ 
\hline
& Linux\_lock(2)				&&6			&&   47 && {\bf 21} &&   {\bf 21}					&&  0.03s  &&  0.02s &&  {\bf 0.01s}							&\\
& Linux\_lock(3)				&&6			&&   14 187 799 && {\bf 412 814} &&   {\bf 412 814}	&&  16m01s  &&  36.36s &&  {\bf 33.21s}						&\\  \hline
& Ticket\_mutex\_c11(3)			&& 10		&&   8 054 && {\bf 4 026} &&   {\bf 4 026}			&&  0.69s  &&  50.76s &&  {\bf 0.25s}						&\\ 
& Correia\_Ramalhete(3)		&& 5			&&   {\bf 5 355} && {\bf 5 355} &&   {\bf 5 355}		&&  1.10s  &&  {\bf 0.77s} &&  0.92s							&\\ 
\hline
\end{tabular}
\caption{
Comparison of the performance of the three tools  when running on 
concurrent data structure benchmarks. 
The {\it LB} column and the {\it t/o} entry have the same meanings as in Table~\ref{classic-benchs:table}. 
}
\label{data-structure-benchs:table:one}
%\vspace{-5mm}
\end{table*}

Table~\ref{data-structure-benchs:table:one} gives
 the  results  on the  four verification tasks derived from
\texttt{Linux\_lock(N)},
\texttt{Ticket\_mutex\_c11(N)},
 and \texttt{Correia\_Ramalhete(N)} algorithms.
 We observe that both
  \textsc{Tracer} and 
\textsc{Rcmc}
have  good performance on these benchmarks.
\textsc{Tracer} always generates fewer executions
than \textsc{Rcmc} and \textsc{CDSChecker}, which coincides  with
our previous observations
in~\cref{litmus:section}--\cref{more:synthetic:bench:section}.
To be more precise,
\textsc{Tracer} and \textsc{Rcmc} explore 
 the same number of  executions in  the four benchmarks and \textsc{Tracer}   has a slightly better performance than \textsc{Rcmc} in 3 of 4 cases.
 In \texttt{Correia\_Ramalhete(3)},
  the numbers of  executions explored
by   \textsc{CDSChecker}, \textsc{Rcmc}, and \textsc{Tracer}
are exactly the same,
and \textsc{Rcmc} is slightly faster than \textsc{CDSChecker} and \textsc{Tracer}.
Moreover,
when we increase the number $N$  of threads in $\texttt{Linux\_lock(N)}$
from 2 to 3,
\textsc{CDSChecker}  significantly increases the 
number of explored executions and the total running time.
We note that
the reported results on these benchmarks
for
\textsc{Rcmc}
are similar to  
the results published in~\cite{KLSV:popl18}.
The reported results for
\textsc{CDSChecker} are not the same as the ones reported 
in~\cite{NoDe:toplas16}
because
all benchmarks in Table~\ref{data-structure-benchs:table:one} are loop-unrolled
and therefore we need not  use the \texttt{-m} and  \texttt{-y} arguments
to control
memory liveness and thread yield-based fairness
for  \textsc{CDSChecker} as in~\cite{NoDe:toplas16}.

\begin{table*}[tb]
\centering
\small
\begin{tabular}{  rl   rr  rrrrr  rrrrr r}
\hline
& \multirow{2}{*}{{\bf Program}} &\;\;\;\;\;\;& \multirow{2}{*}{{\bf LB}}   	&\;\;\;\;\;\;& \multicolumn{3}{c}{{\bf Executions Explored}} &\;\;\;\;\;\;& \multicolumn{3}{c}{{\bf Total Running Time}} 	&\\ 
\cline{6-8}
\cline{10-12}
&								&& 			&& \textsc{CDSChecker} &\;\;\;\;\;\;& \textsc{Tracer}	&& \textsc{CDSChecker} &\;\;\;\;\;\;& \textsc{Tracer} 		&\\ 
\hline
& Barrier(3)						&& 10		&&   62 649  && {\bf 31 944}					&& 5.45s   &&  {\bf 1.88s}							&\\ 
& Seqlock(3)						&& 5			&&   17 792  && {\bf 14 864}					&& 1.17s   &&  {\bf 0.77s}							&\\ 
& MPMC\_queue(3)					&& 5			&&   621 882  && {\bf 239 254}					&& 2m11s   &&  {\bf 44.50s}						&\\ 
& MCS\_lock(2)						&& 5			&&   92 210  && {\bf 70 072}					&& 8.91s   &&  {\bf 5.19s}								&\\ 
& Cliffc\_hashtable(4)				&& 0			&&   9 520  && {\bf 4 576}						&& 2.68s   &&  {\bf 0.94s}							&\\ 
& Concurrent\_hashmap(4)			&& 0			&&   110  && {\bf 42}							&& 0.01s   &&  {\bf 0.00s}							&\\ 
\hline
& Chase\_Lev\_deque(2) ({\bf deadlock})	&& 0			&&  162 306  && {\bf 20 852}					&& 20.65s   &&  {\bf 2.10s}						&\\
& SPSC\_queue(2)	({\bf deadlock})		&&3			&&  754    	    && {\bf 57}						&& 0.18s    &&  {\bf 0.01s}							&\\
\hline
\end{tabular}
\caption{
Comparison of the performance of   \textsc{CDSChecker}  and  \textsc{Tracer} when running on 
eight concurrent data structure benchmarks. 
The {\it LB} column and the {\it t/o} entry have the same meanings as in Table~\ref{classic-benchs:table}. 
}
\label{data-structure-benchs:table:eight}
%\vspace{-5mm}
\end{table*}

In the following, we show the performance of \textsc{Tracer} and \textsc{CDSChecker} 
on the remaining algorithms.
We exclude \textsc{Rcmc} from the comparison since the tool currently cannot handle these algorithms~\cite{rcmc:dicuss:private}.
As it was done in~\cref{larger:tests:section},
we  let \textsc{Tracer} and \textsc{CDSChecker} 
exhaustively explore all possible executions even in the cases
where they detect some errors.

Table~\ref{data-structure-benchs:table:eight} compares 
\textsc{Tracer} and \textsc{CDSChecker} 
on eight algorithms taken from~\cite{NoDe:toplas16}.
We observe that
both 
 \textsc{Tracer} and \textsc{CDSChecker}  handle these benchmarks quite well. 
In all examples, \textsc{Tracer} generates fewer executions than \textsc{CDSChecker} 
and has a better performance, which is also consistent with 
our  previous observations.
Interestingly,  both tools can detect two deadlock errors in \texttt{Chase\_Lev\_deque(2)}
and \texttt{SPSC\_queue(2)}. 
The former  is due to our weakening 
of \texttt{seq-cst} atomic accesses used in the original benchmark to \texttt{acquire} and \texttt{release} atomic accesses.
The later is a well-known known bug in~\cite{spsc:bug}. 
%
%For deadlock benchmarks, we note that as it was done in  
%%
%Interestingly, \textsc{CDSChecker} explores a fewer number of executions than \textsc{Tracer} in the \texttt{SPSC\_queue} benchmark.
%%
%We conjecture
%that the reason is
%related to the way how 
%\textsc{CDSChecker}
%handles
%deadlocks.
%In fact, when a deadlock is encountered
%by \textsc{CDSChecker},
%the tool may not explore any  other execution
% containing the pattern 
%that has led to the  deadlock
%(even 
%if this execution is not equivalent
%to others).
%This explains why \textsc{CDSChecker} generates
%fewer executions
%than \textsc{Tracer}
%for the \texttt{SPSC\_queue} example. 
%Observe that
% \textsc{Tracer} will generate any non-equivalent
% executions
% even if they have the same deadlock pattern.
%Finally note that
%this does not happen in the case of 
%\texttt{Chase\_Lev\_deque} benchmark.

\begin{table*}[tb]
\centering
\small
\begin{tabular}{  rl   rr  rrrrr  rrrrr r}
\hline
& \multirow{2}{*}{{\bf Program}} &\;& \multirow{2}{*}{{\bf LB}}   	&\;\;\;& \multicolumn{3}{c}{{\bf Executions Explored}} &\;\;\;& \multicolumn{3}{c}{{\bf Total Running Time}} 	&\\ 
\cline{6-8}
\cline{10-12}
&								&& 			&& \textsc{CDSChecker} &\;\;\;\;& \textsc{Tracer}	&& \textsc{CDSChecker} &\;\;\;\;& \textsc{Tracer} 		&\\ 
\hline
&Tidex(3)							&& 5			&&   4 676  && {\bf 748}						&& 0.37s   &&  {\bf 0.04s}							&\\ 
&Tidex\_nps(2)						&& 5			&&   10 257  && {\bf 10 254}					&& 6m15s   &&  {\bf 22.58s}						&\\ 
& CLH\_c11(3)						&& 10		&&   2 562  && {\bf 732}						&& 0.22s   &&  {\bf 0.06s}							&\\ 
&  CLH\_rwlock\_c11(3)				&& 10		&&   20  && {\bf 6}							&& 0.01s   &&  {\bf 0.00s}							&\\ 
& MPSC\_c11(3)					&& 10		&&   12 937  && {\bf 4 824}					&& 1.20s   &&  {\bf 0.36s}							&\\ 
\hline
& Correia\_Ramalhete\_turn(3)	 ({\bf mutex broken}) && 2	&&  441 494  && {\bf 96 184}					&& 1m17s   &&  {\bf 12.78s}						&\\ 
\hline
\end{tabular}
\caption{
Comparison of the performance of   \textsc{CDSChecker}  and  \textsc{Tracer} when running on 
six concurrent data structure benchmarks. 
The {\it LB} column and the {\it t/o} entry have the same meanings as in Table~\ref{classic-benchs:table}. 
}
\label{data-structure-benchs:table:six}
%\vspace{-5mm}
\end{table*}

Table~\ref{data-structure-benchs:table:six} compares 
\textsc{Tracer} and \textsc{CDSChecker} 
on the remaining six algorithms taken from~\cite{c11:bench}.
In a similar way to Table~\ref{data-structure-benchs:table:eight},
both tools  handle the benchmarks quite well. 
Moreover, 
we also observe that
\textsc{Tracer} explores fewer executions than \textsc{CDSChecker} 
and has a better performance.
Interestingly, 
although
both tools generate almost the same number of executions in \texttt{Tidex\_nps(2)},
 \textsc{Tracer} is significantly faster than \textsc{CDSChecker}.
Finally, both tools can efficiently detect a violation of the mutual exclusion property in 
\texttt{Correia\_Ramalhete\_turn(3)}.
In the same way as the two deadlocks in Table~\ref{data-structure-benchs:table:eight},
this violation  
is due to our weakening 
of \texttt{seq-cst} atomic accesses used in the original benchmark to \texttt{acquire} and \texttt{release} atomic accesses.

%
%From Table~\ref{data-structure-benchs:table:eight} and
% Table~\ref{data-structure-benchs:table:six},
%  we also observe that in many cases 
%  it is possible to  rewrite 
%   programs just by using 
%   \texttt{acquire} and \texttt{release} atomic accesses
%   under the RA semantics
%   and still preserve the correctness.
%%
%However, for some program, we  might need to use additional RMW accesses as fences
%to obtain correctness.

\subsection{Conclusions of the Experiments}
As expected, the experiments show that, for many 
benchmarks, the number of weak traces is significantly smaller
than the number of total traces.
On these examples, 
\textsc{Tracer}
(which generates  optimal numbers of executions w.r.t.\ weak traces),
is  much more efficient than
\textsc{CDSChecker} and
\textsc{Rcmc}.
The results
also show that
\textsc{Tracer}
has better performance
and scales better
to more extensive programs, even on benchmarks
where the numbers of total and weak traces 
are equal, in which case
\textsc{Tracer}
explores the same number of executions
as the other tools.

%The reason is that 
%\textsc{CDSChecker} is able to detect
%deadlocks using shorter
%executions
%than \textsc{Tracer}
%which its explored deadlock execution
%contains more instructions
%than needed. 
%Observe that
%this result
%that the number of executions
%explored by \textsc{Tracer}
%is larger than \textsc{CDSChecker}
%due to
%the fact that Tracer explores
%more non-relevant executions
%
% 
%
%the benchmark contains a deadlock error and due to  different schedulers, \textsc{CDSChecker} can earlier detect deadlock errors in executed executions (this is not the case in \texttt{chase\_lev\_deque}),
%pruning the search space sooner and resulting in a fewer number of executions.

%The reason is that 
%the benchmark contains a deadlock error and  \textsc{CDSChecker} stops  the exploration as soon as it detects the deadlock errors (this is not the case in \texttt{chase\_lev\_deque}),
%pruning the search space sooner and resulting in a fewer number of executions.

%\phongtodo{}

\section{Related Work}
\label{related:section}

Since the pioneering work of Verisoft~\cite{Godefroid:popl97,Godefroid:verisoft-journal} and CHESS~\cite{MQBBNN:chess}, stateless model checking (SMC), coupled with (dynamic) partial order
reduction techniques (e.g.~\cite{FG:dpor,abdulla2014optimal,RodriguezSSK15})
has been successfully applied to real life
programs~\cite{GoHaJa:heartbeat,KoSa:spin17}.
The method of~\cite{abdulla2014optimal}
is optimal
w.r.t.\ Mazurkiewicz traces (i.e., Shasha-Snir traces) under SC
while 
our algorithm is designed for RA and it is optimal w.r.t.\ weak traces.
With modifications, SMC techniques have been subsequently applied
to the TSO and PSO weak memory
models~\cite{tacas15:tso,DBLP:conf/pldi/ZhangKW15,DBLP:conf/oopsla/DemskyL15}, and
POWER~\cite{DBLP:conf/cav/AbdullaAJL16}.
%, e.g., in Nidhugg~\cite{tacas15:tso}.
Common to all these approaches is
that
they explore at least one execution for each Mazurkiewicz~\cite{Mazurkiewicz:traces} or Shasha-Snir~\cite{ShSn:parallel} traces.
This induces a limit on the efficiency of the corresponding tools.

Several recent DPOR techniques try to exploit the potential offered by
a weaker equivalence
than Mazurkiewicz and Shasha-Snir traces~\cite{NoDe:toplas16,DBLP:conf/pldi/Huang15,DBLP:conf/oopsla/Huang016,DC-DPOR@POPL-18}.
Maximal causality reduction (MCR)
 is a technique based on exploring the possible {\em values} that reads can see, instead
of the possible value-producing writes, as in our approach. MCR has been developed
for SC~\cite{DBLP:conf/pldi/Huang15} and 
TSO and PSO~\cite{DBLP:conf/oopsla/Huang016}. 
It may in some cases explore
fewer traces than our approach, but it 
relies on potentially costly calls to an
SMT solver to find new executions. 
%
%In practise, MCR may still explore a significant number of different executions
%with the same values for the reads.
In practice, MCR~\cite{DBLP:conf/pldi/Huang15} may not be optimal w.r.t. its partitioning (see~\cite{DC-DPOR@POPL-18}) while our algorithm is provably optimal w.r.t. weak traces. 
Moreover, it remains to be seen whether MCR can be
adapted to the RA semantics, and how it would compare.
\citet{DC-DPOR@POPL-18}
proposes
a DPOR algorithm using  a similar equivalence (that is based on $\po$ and $\rf$ relations)
as our weak trace but under SC.
Except for the minimal case of an acyclic communication graph~\cite{DC-DPOR@POPL-18},
they may still explore a significant number of different executions
with the same $\rf$ relation.
Furthermore, checking the consistency of a trace
under SC is an NP-complete problem~\cite{DC-DPOR@POPL-18}. 
As shown in our results, checking
the consistency of a trace under the RA semantics can be efficiently done  by 
polynomial time algorithms (cf.\ \cref{saturated:section}).

Recently, SMC was also adapted to (variants of) the
C/C++11 memory model, which includes RA,
producing the tools \textsc{CDSChecker}~\cite{NoDe:toplas16} and \textsc{Rcmc}~\cite{KLSV:popl18}.
\textsc{CDSChecker}  maintains a coherence order that need not be total, but not in an
optimal way. It can generate inconsistent executions, which must afterward
be validated.
\textsc{Rcmc} has two options: \textsc{Rc11} and \textsc{Wrc11}.
Under \textsc{Rc11}, it maintains only  total coherence orders;
under \textsc{Wrc11} it does not, which may 
generate RA-inconsistent executions (which are then not validated).
\textsc{Rcmc} is  optimal under the criterion of total coherence order, but
only in the absence of RMW operations. In contrast, our technique is provably
optimal w.r.t.\ weak traces, and including RMWs. On the other hand,
\textsc{CDSChecker} and \textsc{Rcmc} also cover the different access modes of the C/C++11 memory model.

Bounded model checking  can adapt to various weak memory models (e.g.~\cite{AlKNT13,AlglaveKT13,TorlakVD10}). There is no report on using them
for RA, but experiments for POWER~\cite{DBLP:conf/cav/AbdullaAJL16} concluded
that Nidhugg is at least as efficient as~\cite{AlKNT13}.

Beyond SMC techniques for weak memory models, there have been many works related to the verification of programs running under weak memory models  (e.g.,~\cite{LNPVY12,kupersteinvy:11,BAM07,AtigBBM10,p2-1,p3-1,p1}).
%% Some of these works propose precise analysis techniques for finite-state programs under relaxed memory models (e.g.,~\cite{abdulla2012counter,BouajjaniDM13,DM14}).
Some works propose algorithms and tools for monitoring and testing programs running under weak memory models (e.g.,~\cite{BM08,BurSS11,LNPVY12}).

\section{Conclusions and Future Work}
\label{conclusions:section}

We have presented a new approach to defining
DPOR algorithms which is optimal in the sense that
it generates at most one trace with a given
program order and read-from relation.
%\patodo{program-order or program order. Same for read-from.
%Be consistent!}
%
We have instantiated the approach  for the RA fragment of C/C++11.
Our tool  demonstrates that our method is substantially
more efficient than state-of-the-art tools that handle the same 
fragment.

Although  we only consider the RA semantics in this paper, we
believe that our approach is general and can be extended to other 
memory models.
For instance, we can extend the approach to the SRA (Strong RA) semantics
\cite{DBLP:conf/popl/LahavGV16} by a  modification of the
sets of readable and visible events (cf.\ \cref{saturated:section}).
It is interesting to see whether we can also handle the {\it relaxed} fragment of C/C++11 by employing
a swapping mechanism for event speculations similar to the one we 
have proposed in this paper  for treating postponed write events
(cf.\ \cref{dpor:section}).
For other models such as SC, our saturation scheme is necessarily
not complete, since saturation for such models amounts
to solving an NP-complete problem \cite{DC-DPOR@POPL-18}.
However, we believe that by
maintaining saturated traces  and only
running the costly operations mentioned in \cite{DC-DPOR@POPL-18}
``by demand'', we can  substantially improve efficiency
even under the SC semantics.

\begin{acks}                 
We thank Brian Demsky, Peizhao Ou, Konstantinos Sagonas, Michalis Kokologiannakis, Carl Leonardsson, Magnus Lång, and the OOPSLA'18 reviewers for their helpful feedback. 
This work 
was carried out within the UPMARC Linnaeus centre of excellence
(Uppsala Programming for Multicore Architectures
Research Center). 
\end{acks}

\bibliography{bibdatabase.bib}

\end{document}